%% file: ttbb_PRD_rev.tex
\documentclass[11pt,epsf]{article}
 \usepackage{amsmath}
 \usepackage{graphicx}
 \usepackage[merge,numbers,sort&compress]{natbib}
 \usepackage[T1]{fontenc}
 \usepackage{booktabs}
 \usepackage{xcolor} 
 \usepackage{xspace}
 \usepackage{dcolumn}
 \usepackage{placeins}
 \usepackage[colorlinks=true,citecolor=blue!50!black,linkcolor=black]{hyperref}
 \usepackage{caption}
 \usepackage
 [subrefformat=parens,position=top,skip=-15pt,margin=15pt,justification=justified,singlelinecheck=false]
 {subcaption}

\numberwithin{equation}{section}

\input{macros}

\begin{document}

\title{\hfill ~\\[-30mm]
\phantom{h} \hfill\mbox{\small Cavendish-HEP-20/06, ZU-TH 27/20}
\\[1cm]
\vspace{13mm}   \textbf{Full NLO QCD corrections to off-shell ttbb production}}

\date{}
\author{
Ansgar Denner$^{1\,}$\footnote{E-mail:
  \texttt{ansgar.denner@physik.uni-wuerzburg.de}},
Jean-Nicolas Lang$^{2\,}$\footnote{E-mail:
\texttt{jlang@physik.uzh.ch}},
Mathieu Pellen$^{3\,}$\footnote{E-mail:
  \texttt{mpellen@hep.phy.cam.ac.uk}},
\\[9mm]
{\small\it
$^1$ Universit\"at W\"urzburg, %
        Institut f\"ur Theoretische Physik und Astrophysik,} \\ %
{\small\it Emil-Hilb-Weg 22, \linebreak %
        97074 W\"urzburg, %
        Germany}\\[3mm]
{\small\it
$^2$ Universit\"at Z\"urich, Physik-Institut,} \\ %
{\small\it CH-8057 Z\"urich, Switzerland}\\[3mm]
{\small\it
$^3$ University of Cambridge, Cavendish Laboratory,} \\ %
{\small\it 19 JJ Thomson Avenue, Cambridge CB3 0HE, United Kingdom}\\[3mm]
        }
\maketitle

\typeout{\the\textwidth}
\typeout{\the\linewidth}


\begin{abstract}
  \noindent\sloppy
  In this article, we report on the computation of the NLO QCD
  corrections to $\Pp\Pp \to
  \mu^-\bar\nu_{\mu}\Pe^+\nu_{\Pe}\bar\Pb\Pb\bar\Pb\Pb$ at the LHC,
  which is an irreducible background to
  $\Pp\Pp\to\Pt\bar\Pt\PH\bigl(\to\Pb\bar\Pb\bigr)$.
  This is the first time that a full NLO computation for a $2\to8$
  process with 6 external strongly-interacting partons is made public.
  No approximations are used, and all off-shell and interference
  effects are taken into account.  Cross sections and differential
  distributions from the full computation are compared to results
  obtained by using a double-pole approximation for the top quarks.
  The difference between the full calculation and the one using the
  double-pole approximation is in
  general below 5\% but can reach 10\% in some regions of phase space.
\end{abstract}

\typeout{\the\textwidth}
\typeout{\the\linewidth}

\newpage

\section{Introduction}

The physics programme of the Large Hadron Collider (LHC) is driven by
the measurement of fundamental parameters of the Standard Model (SM)
of particle physics.  These range from masses and widths to couplings
of elementary particles.  Such parameters are experimentally measured
using specific physical processes that are particularly sensitive to
them.  Given the complexity of the LHC environment, the sought-after
signal is often polluted by background processes that mimic the final
state of the signal.  Even worse, there are also irreducible
backgrounds that have exactly the same final state as the signal and
differ only in the order of the strong and electroweak couplings.
Thus, the extraction of fundamental parameters requires the
subtraction of contributions of background processes from the
measurements.  Therefore, in order to allow for a precise measurement
of parameters, theoretical predictions with high precision are
required for both the signals and the backgrounds.

A prime example is the extraction of the Higgs coupling to top quarks
from the measurement of $\Pp\Pp\to\Pt\bar\Pt\PH$.  Given the large
branching ratio of the Higgs boson into a pair of bottom--antibottom
quarks, it is one of the favourite channels for the measurement of
$\Pp\Pp\to\Pt\bar\Pt\PH$.  Taking into account the top-quark decay
products, the complete signal process reads $\Pp\Pp \to
\mu^-\bar\nu_{\mu}\Pe^+\nu_{\Pe}\bar\Pb\Pb\bar\Pb\Pb$ at order%
\footnote{Squared Yukawa couplings are understood as order $\alpha$.}
$\order{\alphas^2 \alpha^6 }$.  The same process receives
contributions at order $\order{\alphas^4 \alpha^4}$, where the
bottom--antibottom pair results from {a strong interaction}.  In
recent years, much attention has been devoted to the computation of
the signal
\cite{Beenakker:2001rj,Reina:2001sf,Beenakker:2002nc,Dawson:2003zu,Frederix:2011zi,Garzelli:2011vp,Hartanto:2015uka,Denner:2015yca,Kulesza:2015vda,Broggio:2015lya,Kulesza:2016vnq,Broggio:2016lfj,Denner:2016wet,Broggio:2019ewu,Kulesza:2020nfh,Frixione:2014qaa,Yu:2014cka,Frixione:2015zaa,Badger:2016bpw}
as well as the background process
\cite{Bredenstein:2008zb,Bredenstein:2009aj,Bevilacqua:2009zn,Bredenstein:2010rs,Cascioli:2013era,Kardos:2013vxa,Garzelli:2014aba,Bevilacqua:2017cru,Jezo:2018yaf,Buccioni:2019plc}.
In particular, it has been found that theoretical predictions for the
background can vary substantially depending on the exact matching
and/or parton shower used and tend to overestimate the experimental
measurement by $30$--$50\%$
\cite{deFlorian:2016spz,talkPozzoriniTOP2018,talkSiegertZPW2020}.  In
such predictions, the process is computed with on-shell top quarks,
\ie $\Pp\Pp\to\Pt\bar\Pt\Pb\bar\Pb$, which are subsequently decayed by
a parton-shower program.  However, top quarks also generate bottom
quarks while decaying.  Therefore, the physically relevant
irreducible-background process is $\Pp\Pp \to
\mu^-\bar\nu_{\mu}\Pe^+\nu_{\Pe}\bar\Pb\Pb\bar\Pb\Pb$ at order
$\order{\alphas^4 \alpha^4}$.  The reason why studies have so far
focussed on an on-shell description of the top quark is the complexity
of the above process \cite{Denner:2014wka}.  Indeed, it is a $2\to8$
process with 6 external strongly-interacting particles and multiple
intermediate resonances. Such a complex process has never
been computed at next-to-leading order (NLO) QCD accuracy.%
\footnote{In \citeres{Bevilacqua:2019cvp,Bevilacqua:2020pzy,Denner:2020hgg}, $2\to8$
  computations at NLO have been presented with 4 external
  strongly-interacting partons.  The calculation of
  \citere{Anger:2017glm}, on the other hand, involves up to 7 external
  QCD particles but for a $2\to7$ process.}

Experimentally, the cross section for $\Pp\Pp\to\Pt\bar\Pt\Pb\bar\Pb$
has been measured by the ATLAS and CMS collaborations
\cite{Aaboud:2018eki,Sirunyan:2020kga}.  The production of a Higgs
boson in association with a top--antitop pair was observed by both
ATLAS and CMS by combining various Higgs decay channels
\cite{Aaboud:2018urx,Sirunyan:2018hoz}.  For the specific channel with
the Higgs decaying into a bottom--antibottom pair searches have been
performed as well \cite{Aaboud:2017rss,Sirunyan:2018mvw}.

In this article we report on the computation of the NLO QCD
\sloppy
corrections to $\Pp\Pp \to
\mu^-\bar\nu_{\mu}\Pe^+\nu_{\Pe}\bar\Pb\Pb\bar\Pb\Pb$ at order
$\order{\alphas^5 \alpha^4}$ at the LHC.  No
approximations are used, and all off-shell as well as all interference
effects are taken into account.  This computation has been made
possible by the use of the efficient Monte Carlo integrator \mocanlo
in combination with \recolatwo
\cite{Actis:2012qn,Actis:2016mpe,Denner:2017vms,Denner:2017wsf} in association with \otter
\cite{otter:2020}, a new tensor integral library, and \collier
\cite{Denner:2014gla,Denner:2016kdg}.
In addition to the full computation, a calculation using a double-pole
approximation (DPA) for the virtual corrections that retains only contributions with top
  and antitop quarks decaying into a lepton--neutrino pair and a
  bottom quark,
has been performed \cite{Denner:2000bj}.
Comparison of the DPA results with those of the full calculation serves as a consistency check and provides an indication of contributions 
beyond the approximation of on-shell top quarks. The results are
presented in the form of cross sections and differential
distributions.  We emphasise that the present computations certainly
do not answer all questions regarding the theoretical description of
$\Pt\bar\Pt\Pb\bar\Pb$ on its own.  Nonetheless, they constitute an
important piece of information that could serve as a basis for future
comparative studies.

The article is split into two main parts.  Section
\ref{se:details_computation} describes the computations carried out, while
\refse{se:results} focuses on the presentation of the numerical
results.  A summary of the main findings of the present work is
provided in \refse{se:conclusion}.

\section{Details of the calculations}
\label{se:details_computation}

\subsection{Process definition}

The hadronic process under investigation is the production of a
top--antitop pair in association with a bottom--antibottom pair at the
LHC.  Considering the leptonic decays of the top quarks, the process
reads
\begin{equation} \label{eq:LOprocess}
 \Pp\Pp \to \mu^-\bar\nu_{\mu}\Pe^+\nu_{\Pe}\bar\Pb\Pb\bar\Pb\Pb+X.
\end{equation}
At leading order (LO), the dominant contribution is of order
$\order{\alphas^4 \alpha^4}$.  The process \refeq{eq:LOprocess}
constitutes the irreducible-background to
$\Pp\Pp\to\Pt\bar\Pt\PH\bigl(\to \Pb\bar\Pb \bigr)$, which is of order
$\order{\alphas^2 \alpha^6}$. The partonic processes contributing to
hadronic events have two gluons, a quark--antiquark pair, and two
bottom quarks or two antibottom quarks in the
initial state,
\begin{align}
\label{eq:bornprocesses}
 \Pg\Pg &\to \mu^-\bar\nu_{\mu}\Pe^+\nu_{\Pe}\bar\Pb\Pb\bar\Pb\Pb, \notag \\
 q \bar q/ \bar q q &\to \mu^-\bar\nu_{\mu}\Pe^+\nu_{\Pe}\bar\Pb\Pb\bar\Pb\Pb, \notag \\
 \Pb \bar \Pb / \bar \Pb \Pb &\to \mu^-\bar\nu_{\mu}\Pe^+\nu_{\Pe}\bar\Pb\Pb\bar\Pb\Pb,\notag\\
\Pb\Pb &\to \mu^-\bar\nu_{\mu}\Pe^+\nu_{\Pe}\bar\Pb\Pb\Pb\Pb,\notag\\
\bar\Pb\bar\Pb &\to \mu^-\bar\nu_{\mu}\Pe^+\nu_{\Pe}\bar\Pb\Pb\bar\Pb\bar\Pb,
\end{align}
with $q=\Pu,\Pd,\Pc,\Ps$.  For the $\Pg\Pg$ channel, $3904$ Feynman
diagrams contribute at LO, while for the $q \bar q$ ones there are
$930$. The channels with bottom quarks in the initial state furnish
$2790$ Feynman diagrams each.

The NLO QCD corrections to the LO process of order $\order{\alphas^4
  \alpha^4}$ are thus defined at order $\order{\alphas^5 \alpha^4}$
and include real and virtual contributions.  The real NLO corrections
are obtained upon adding an extra real gluon in the final state of the
processes \refeq{eq:bornprocesses} and by possibly crossing this extra
gluon and one of the initial-state partons.  Consequently, the
relevant processes for the real NLO corrections read:
\begin{align}
\label{eq:realprocesses}
 \Pg\Pg &\to \mu^-\bar\nu_{\mu}\Pe^+\nu_{\Pe}\bar\Pb\Pb\bar\Pb\Pb\Pg, \notag \\
 q \bar q/ \bar q q &\to \mu^-\bar\nu_{\mu}\Pe^+\nu_{\Pe}\bar\Pb\Pb\bar\Pb\Pb\Pg, \notag \\
 \Pg \bar q/ \bar q \Pg &\to \mu^-\bar\nu_{\mu}\Pe^+\nu_{\Pe}\bar\Pb\Pb\bar\Pb\Pb\bar\Pq, \notag \\
 \Pg q/ q \Pg &\to \mu^-\bar\nu_{\mu}\Pe^+\nu_{\Pe}\bar\Pb\Pb\bar\Pb\Pb\Pq, \notag \\
 \Pb \bar \Pb / \bar \Pb \Pb &\to \mu^-\bar\nu_{\mu}\Pe^+\nu_{\Pe}\bar\Pb\Pb\bar\Pb\Pb\Pg, \notag \\
 \Pg \bar \Pb / \bar \Pb \Pg &\to \mu^-\bar\nu_{\mu}\Pe^+\nu_{\Pe}\bar\Pb\Pb\bar\Pb\Pb\bar\Pb, \notag \\
 \Pg \Pb / \Pb \Pg &\to \mu^-\bar\nu_{\mu}\Pe^+\nu_{\Pe}\bar\Pb\Pb\bar\Pb\Pb\Pb,\notag\\
\Pb\Pb &\to \mu^-\bar\nu_{\mu}\Pe^+\nu_{\Pe}\bar\Pb\Pb\Pb\Pb\Pg,\notag\\
\bar\Pb\bar\Pb &\to \mu^-\bar\nu_{\mu}\Pe^+\nu_{\Pe}\bar\Pb\Pb\bar\Pb\bar\Pb\Pg.
\end{align}

On the other hand, the virtual corrections are made of one-loop
amplitudes interfered with tree-level ones.  The one-loop diagrams are
built from the tree-level ones by inserting a virtual gluon and closed
quark loops in all possible ways.  Note that here no mixed QCD--EW
corrections are present at this order as it can be the case in other
computations for top--antitop production
\cite{Denner:2016jyo,Denner:2016wet,Denner:2017kzu}.
For illustration, the one-loop virtual amplitude of the $\Pg\Pg$ channel
involves more than $200 000$  Feynman diagrams, while the
corresponding real tree-level one possesses $41364$ diagrams. Moreover, the
virtual corrections to the $\Pg\Pg$ channel feature up to rank-6 8-point
integrals, involve more than 10000 different tensor integrals, and evaluate in
2.6 seconds per phase space point on average on a
Intel(R) Core(TM) i7-7700 CPU @ 3.60GHz.

\subsection{Description of the computations}
\label{se:description_computation}

\subsubsection*{Computation based on complete NLO matrix elements}

\begin{figure}
\captionsetup{skip=0pt}
\centering
\begin{subfigure}[b]{0.3\textwidth}
\centering
\includegraphics[page=1,scale=0.9]{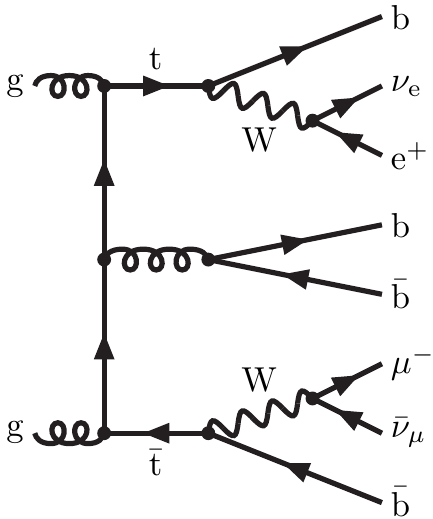}
\caption{Doubly top resonant}
\label{fig:diag_2resbb}
\end{subfigure}
\begin{subfigure}[b]{0.37\textwidth}
\centering
\includegraphics[page=2,scale=0.9]{diagrams.pdf}
\caption{Doubly top resonant}
\label{fig:diag_2restt}
\end{subfigure}%
\begin{subfigure}[b]{0.3\textwidth}
\centering
\includegraphics[page=3,scale=0.9]{diagrams.pdf}
\caption{Doubly top resonant}
\label{fig:diag_2restb}
\end{subfigure}%
\\[6ex]
\centering
\begin{subfigure}[b]{0.45\textwidth}
\centering
\includegraphics[page=7,scale=0.9]{diagrams.pdf}
\subcaption{Three top/antitop propagators}
\label{fig:diag_ttg}
\end{subfigure}
\begin{subfigure}[b]{0.45\textwidth}
\centering
\includegraphics[page=8,scale=0.9]{diagrams.pdf}
\subcaption{$\Pt\bar\Pt$ intermediate state}
\label{fig:diag_twbg}
\end{subfigure}%
\\[6ex]
\centering
\begin{subfigure}[b]{0.3\textwidth}
\centering
\includegraphics[page=4,scale=0.9]{diagrams.pdf}
\subcaption{Singly top resonant}
\label{fig:diag_1resbb}
\end{subfigure}
\begin{subfigure}[b]{0.37\textwidth}
\centering
\includegraphics[page=5,scale=0.9]{diagrams.pdf}
\subcaption{Singly top resonant}
\label{fig:diag_1restb}
\end{subfigure}%
\begin{subfigure}[b]{0.3\textwidth}
\centering
\includegraphics[page=6,scale=0.9]{diagrams.pdf}
\subcaption{Non top resonant}
\label{fig:diag_0resbb}
\end{subfigure}
\\[6ex]
\caption{Sample LO diagrams for the partonic channel $\Pg\Pg \to
  \mu^-\bar\nu_{\mu}\Pe^+\nu_{\Pe}\bar\Pb\Pb\bar\Pb\Pb$.}
\label{fig:diagrams}
\end{figure}

The full computation comprises all possible real and virtual
corrections mentioned above that contribute to the cross section at order
$\order{\alphas^5 \alpha^4}$, \ie all partonic channels and all Feynman diagrams of order
  $\order{g_\mathrm{s}^5e^4}$, contributing to the cross section in
  the order $\order{\alphas^5 \alpha^4}$, are taken into account. In
particular, no approximations are employed, and all off-shell as well
as all interference effects are included.  Some of the contributing
diagrams for the partonic channel $\Pg\Pg \to
\mu^-\bar\nu_{\mu}\Pe^+\nu_{\Pe}\bar\Pb\Pb\bar\Pb\Pb$ are shown in
\reffi{fig:diagrams}.  These include diagrams with two top resonances
(\reffis{fig:diag_2resbb}, \ref{fig:diag_2restt},
\ref{fig:diag_2restb}, \ref{fig:diag_twbg}), with three
potential top resonances (\reffis{fig:diag_ttg}), with one
top resonance (\reffis{fig:diag_1resbb}, \ref{fig:diag_1restb}), and
with no top resonance (\reffi{fig:diag_0resbb}).
  The diagram in \reffi{fig:diag_ttg} contains one antitop and
  two top propagators. While both top propagators cannot be
  simultaneously resonant, each one becomes resonant in some part of
  phase space, corresponding to different on-shell processes, \ie
  $\Pt\bar\Pt$ production with the subsequent decays
  $\Pt\to\nu_\Pe\Pe^+\Pb\bar\Pb\Pb$ and
  $\bar\Pt\to\bar\nu_\Pe\Pe^-\bar\Pb$ and $\bar\Pt\Pt\bar\Pb\Pb$
  production with the subsequent decays $\Pt\to\nu_\Pe\Pe^+\Pb$ and
  $\bar\Pt\to\bar\nu_\mu\mu^-\bar\Pb$. On the other hand, the diagram
  in \reffi{fig:diag_twbg} contributes only to $\Pt\bar\Pt$ production
  but not to $\bar\Pt\Pt\bar\Pb\Pb$ production, since there are no
  $\bar\Pt\Pt\bar\Pb\Pb$ intermediate states.

The computation is carried out in the 5-flavour scheme that assumes the
bottom quarks to be massless throughout.
All leptons and quarks
(apart from the top quark) are thus taken to be massless.  Also, all
potentially resonant particles, \ie top quark, W~boson and Z~boson,
are treated within the complex-mass
scheme~\cite{Denner:1999gp,Denner:2005fg,Denner:2019vbn}, ensuring
gauge invariance of all the amplitudes.

\subsubsection*{Double-pole approximation}
\label{se:DPA}

Similar to \citeres{Denner:2016jyo,Denner:2016wet}, in addition to the
full computation we also perform a calculation based on a DPA.  Note
that the DPA is only applied to matrix elements, while the physical
observables are calculated with off-shell kinematics.
Specifically, we examine the \emph{tt-DPA} which consists in retaining
only contributions that feature two resonant top quarks and projecting
the top-quark momenta on shell, apart from those in the denominators
of the resonant propagators, which are kept off shell.  At LO, the
tt-DPA is based on the doubly-top-resonant contributions in the
Born matrix element.  More precisely, we include only those
  Feynman diagrams that contain both the decays $\Pt\to\nu_\Pe\Pe^+\Pb$ and
  $\bar\Pt\to\bar\nu_\mu\mu^-\bar\Pb$ as sub-diagrams, like those in
  \reffis{fig:diag_2resbb}, \ref{fig:diag_2restt},
  \ref{fig:diag_2restb}, and \ref{fig:diag_ttg}, but no diagrams with
  different top decays like the one in \reffi{fig:diag_twbg}.  Moreover, only one
  of the bottoms, say bottom 1, is allowed as decay product of the top
  quark, while the other one, bottom 2, is not, \ie diagrams with
  bottoms interchanged are not contained in the DPA. The same applies
  to the antibottoms with respect to the antitops.  In addition, the
  resulting squared amplitudes from \recola are multiplied by a factor
  4 to ensure the correct symmetry factors.%
\footnote{For $\Pb\Pb$ or $\bar\Pb\bar\Pb$ initial states, the \recola
  amplitudes must be  multiplied by a factor
  3 instead.}   The on-shell projection
  is performed in the same way as described for $\Pt\bar\Pt$
  production in \citere{Denner:2016jyo}. We note that some of the
  diagrams contributing to the DPA, \eg the one in \reffi{fig:diag_ttg}, contain besides
  contributions to  $\Pt\bar\Pt\Pb\bar\Pb$ production also 
  doubly resonant contributions to $\Pt\bar\Pt$
  production with the top (or antitop) decaying into
  $\Pt\to\nu_\Pe\Pe^+\Pb\bar\Pb\Pb$ (or
  $\bar\Pt\to\bar\nu_\mu\mu^-\bar\Pb\bar\Pb\Pb$). Diagrams like the
  one in \reffi{fig:diag_twbg} that  contain $\Pt\bar\Pt$ production
  but not $\Pt\bar\Pt\Pb\bar\Pb$ production as a
  subprocess are not included in the DPA.
As opposed to a narrow-width
  approximation, in the DPA full spin correlations, off-shell
  propagators, as well as the full phase space are taken into account.
  In the tt-DPA calculation, we treat W and Z~bosons in the
  complex-mass scheme.

At NLO the DPA is applied only to the virtual corrections, and also
the doubly-resonant non-factorisable corrections following the
algorithm of
\citeres{Denner:1997ia,Accomando:2004de,Dittmaier:2015bfe} transferred to
QCD are included.  All other contributions of orders
$\order{\alphas^4\alpha^4}$ and $\order{\alphas^5 \alpha^4}$, \ie LO,
real and subtraction terms, are kept exact.

Note that, as in the original DPA computations \cite{Denner:2000bj}, in the
past computations with \mocanlo
\cite{Denner:2016jyo,Denner:2016wet,Biedermann:2016yds} the DPA
(retaining resonant contributions and applying the on-shell projection)
has also been applied to the I-operator in the integrated dipoles.  It
has been noticed \cite{Denner:2020bcz} that when done in combination with small
$\alpha_{\mathrm{dipole}}$ parameter~\cite{Nagy:1998bb}, this tends to
worsen the agreement with the full computation, as it treats large
contributions in the subtracted and re-added real corrections that
should cancel differently. Applying instead the DPA only to the
virtual corrections, with IR singularities subtracted via an
appropriate choice of regularisation parameters,%
\footnote{In practice we discard the IR poles and set the parameter
  {\tt muir} of {\sc Collier} equal to the top-quark mass.}
avoids this mismatch
and provides better agreement with the full calculation.

  The DPA is constructed as a check of the full calculation and
  to provide a good approximation thereof. While a comparison of this
  approximation with the full calculation cannot yield quantitative
  results on off-shell top-quark effects, it nevertheless gives an
  indication on their size. In practice, the actual off-shell effects
  are often even larger, in particular, in specific regions of phase
  space.

\subsubsection*{Tools}

The numerical integration has been carried out with the help of the
multi-channel Monte Carlo integration program \mocanlo.  This code was
developed for the integration of high-multiplicity processes
involving top--antitop pairs and has proven to be particularly
efficient for those and related processes
\cite{Denner:2015yca,Denner:2016jyo,Denner:2016wet,Denner:2017kzu}.
It relies on a multi-channel phase-space integration following
\citeres{Berends:1994pv,Denner:1999gp,Dittmaier:2002ap}.

All one-loop amplitudes in the 8-body phase space have been obtained
from the matrix-element generator
\recolatwo~\cite{Actis:2012qn,Actis:2016mpe,Denner:2017vms,Denner:2017wsf}
in combination with the \otter library that is based on the {\em
  on-the-fly reduction} \cite{Buccioni:2017yxi} and uses the stability
improvements for hard kinematics described in
\citere{Buccioni:2019sur}.  By default, \otter\ uses double-precision
scalar integrals provided by \collier
\cite{Denner:2014gla,Denner:2016kdg} and for exceptional phase-space
points makes targeted use of multi-precision scalar integrals provided
by {\sc OneLOop} \cite{vanHameren:2010cp}. The computation of the
virtual amplitudes has been carried out as well exclusively with the
{\sc COLI} branch of the \collier library, yielding perfect agreement.
The infrared (IR) singularities in the real and virtual corrections
are handled via the Catani--Seymour dipole subtraction formalism
\cite{Catani:1996vz,Nagy:1998bb}. We note that the partonic process
$\Pb\Pg\to \mu^-\bar\nu_{\mu}\Pe^+\nu_{\Pe}\bar\Pb\Pb\bar\Pb\Pb\Pb$
involves 40 Catani--Seymour dipoles, while $\Pg\Pg \to
\mu^-\bar\nu_{\mu}\Pe^+\nu_{\Pe}\bar\Pb\Pb\bar\Pb\Pb\Pg$ involves 30.

\subsubsection*{Validations}

This computation builds on several previous computations for processes
involving top--antitop pairs with \mocanlo
\cite{Denner:2015yca,Denner:2016jyo,Denner:2016wet,Denner:2017kzu,Denner:2020hgg}
which have themselves been thoroughly checked. Within the
dipole-subtraction scheme, the variation of the
$\alpha_{\mathrm{dipole}}$ parameter~\cite{Nagy:1998bb} that narrows
the phase space to singular regions has been used.  For representative
channels a comparison of results for $\alpha_{\mathrm{dipole}}=1$ and
$\alpha_{\mathrm{dipole}}=10^{-2}$ has revealed perfect agreement
within statistical errors.  The results presented below have been
obtained using $\alpha_{\mathrm{dipole}}=10^{-2}$.  Furthermore,
independence on the IR-regulator parameter has been verified for
representative channels, proving IR finiteness.  Finally, the virtual
corrections were computed with \recolatwo both in the conventional 't~Hooft--Feynman
gauge and within the Background-Field method using the two independent
integral-reduction libraries \otter and \collier.  Moreover, for the
gluon-initiated process we verified that when replacing one of
the gluon polarisation vectors at a time by its normalised four-momentum via
$\epsilon^\mu_\Pg\to p_\Pg^\mu/p_\Pg^0$ only in the virtual amplitude, the
corresponding contribution to the cross section integrates to a
numerical zero at the relative level of $10^{-8}$.
Finally, the calculation based on the DPA for the virtual corrections provides a further validation of the full NLO calculation.

\subsection{Input parameters and event selection}

\subsubsection*{Input parameters}

The theoretical predictions presented here are for the LHC at $13\TeV$
centre-of-mass energy.  The on-shell values for the masses and
widths of the gauge bosons \cite{Tanabashi:2018oca},
\begin{equation}
\begin{array}[b]{rcl@{\qquad}rcl}
  \MW^{\rm os} &=& 80.379  \GeV, & \GW^{\rm os} &=& 2.085 \GeV, \\
  \MZ^{\rm os} &=& 91.1876 \GeV, & \GZ^{\rm os} &=& 2.4952\GeV,
\end{array}
\end{equation}
are converted into pole masses according to \cite{Bardin:1988xt}
\begin{eqnarray}
&& M_V = M_V^{\rm os}/c_V, \qquad \Gamma_V = \Gamma_V^{\rm os}/c_V, \notag\\
&& c_V=\sqrt{1+(\Gamma_V^{\rm os}/M_V^{\rm os})^2}, \quad V=\PW,\PZ .
\end{eqnarray}
The latter are used in the calculation.
The top-quark mass and widths are fixed to
\begin{equation}\label{eq:mt}
\Mt = 173\GeV, \qquad \Gt^{\rm LO} = 1.443303\GeV, \qquad \Gt^{\rm NLO} = 1.3444367445\GeV .
\end{equation}
The top-quark width at LO has been computed based on the formulas of
\citere{Jezabek:1988iv}, while the NLO QCD value has been obtained upon
applying the relative QCD corrections of \citere{Basso:2015gca} to the
LO width.  The LO top width is utilised for the LO computation, while the
NLO one is employed in the NLO calculation (including the Born
contributions).

Concerning the electromagnetic coupling $\alpha$, the $G_\mu$ scheme
is applied, where $\alpha$ is fixed from the Fermi constant,
\begin{equation}
\alpha_{G_\mu} = \frac{\sqrt{2}}{\pi}G_\mu\MW^2\left(1-\frac{\MW^2}{\MZ^2}\right) ,
\end{equation}
with
\begin{equation}
G_\mu= 1.16638\times 10^{-5} \GeV^{-2}.
\end{equation}

The sets of parton distribution functions (PDF) NNPDF31 LO and NNPDF31
NLO with $\alphas=0.118$ \cite{Ball:2017nwa} have been used at
LO and NLO, respectively.  The values of $\alphas$ for the dynamical
scales have been taken from the PDF sets which are interfaced through
LHAPDF~\cite{Andersen:2014efa,Buckley:2014ana}. Accordingly, a
variable-flavour-number scheme with at most 5 flavours is used for the
running of~$\alphas$.

The renormalisation and factorisation scales, $\mu_{\mathrm{ren}}$ and
$\mu_{\mathrm{fact}}$, are set equal to
\begin{equation}
\label{eq:scale}
 \mu_0 = \frac{1}{2}
\Biggl[\Biggl( p^{\mathrm{miss}}_{\rT} +
    \sum_{i=\Pl,{\rm J}}
    E_{\rT,i}\Biggr)+2\Mt\Biggr]^{1/2}
\Biggl( \sum_{i={\rm J}} E_{\rT,i}\Biggr)^{1/2} ,
\end{equation}
where $p^{\mathrm{miss}}_{\rT}$ is the transverse component of the
vector sum of the two neutrino momenta and \emph{J} denotes all bottom and light jets after jet clustering.%
\footnote{While for events without real radiation passing the cuts
    $J$ involves precisely 4 bottom jets, for real-radiation events
    the fifth (bottom or light) jet is included as well if it is not
    recombined during the jet clustering.}
The transverse energy $E_{\rT,i}$ of the other particles is defined as
$E_{\rT,i}=\sqrt{p_{\rT,i}^2+m_i^2}$, where $m_i^2$ is the
invariant-mass squared of the object considered (which can be a jet resulting from parton recombination and is thus not necessarily zero).  Note that this choice of scale has
the property not to refer explicitly to a top quark, as it has been
done so far in the literature.  While the first factor in
Eq.~\refeq{eq:scale} serves as a proxy for the typical momentum transfer
in the strong couplings of the top quarks, the second one mimics the
one in the couplings of the bottom quarks in the process. The choice
can be viewed as a modification of the renormalisation scales used in
\citeres{Bredenstein:2010rs,Cascioli:2013era,Jezo:2018yaf} avoiding
identification of the top quarks.

\subsubsection*{Event selection}

The event selection is generic and assumes a \emph{resolved} topology
(as opposed to a \emph{boosted} kinematics).  Quarks and gluons are
clustered using the anti-$k_\rT$ algorithm \cite{Cacciari:2008gp} with
a jet-resolution parameter $R=0.4$.  
Concerning the flavour, the recombination rules read,
\begin{itemize}
 \item $\Pj+\Pj\to\Pj$,
 \item $\Pj_{\rm b}+\Pj\to\Pj_{\rm b}$,
 \item $\Pj_{\rm b}+\Pj_{\rm b}\to\Pj$,
\end{itemize}
where the bottom jet $\Pj_{\rm b}$ contains at least one $\Pb$ or $\bar\Pb$ quark,
while $\Pj$ is a light jet.
The last combination is implemented in order to account for a proper
treatment of the singularity originating from gluon splitting into pairs of
bottom--antibottom quarks and effectively leads to the elimination of
events where the bottom and antibottom quarks are collinear and thus recombined into one jet.
For each bottom jet and charged lepton, a cut on its transverse
momentum and its rapidity is applied,
\begin{align}
 p_{\rT,\Pj_\Pb} >{}& 25\GeV, \qquad |y_{\Pj_\Pb}| < 2.5,\notag\\
 p_{\rT,\Pl} >{}& 20\GeV, \qquad |y_{\Pl}| < 2.5,
\label{eq:cut}
\end{align}
where $\Pl=\Pe^+,\mu^-$. Finally, we require at least 4
  bottom jets, a positron, and a muon passing all these cuts in the final
state.

\section{Numerical results}
\label{se:results}

\subsection{Cross sections}

\begin{figure}
        \center
        \includegraphics[width=0.5\textwidth]{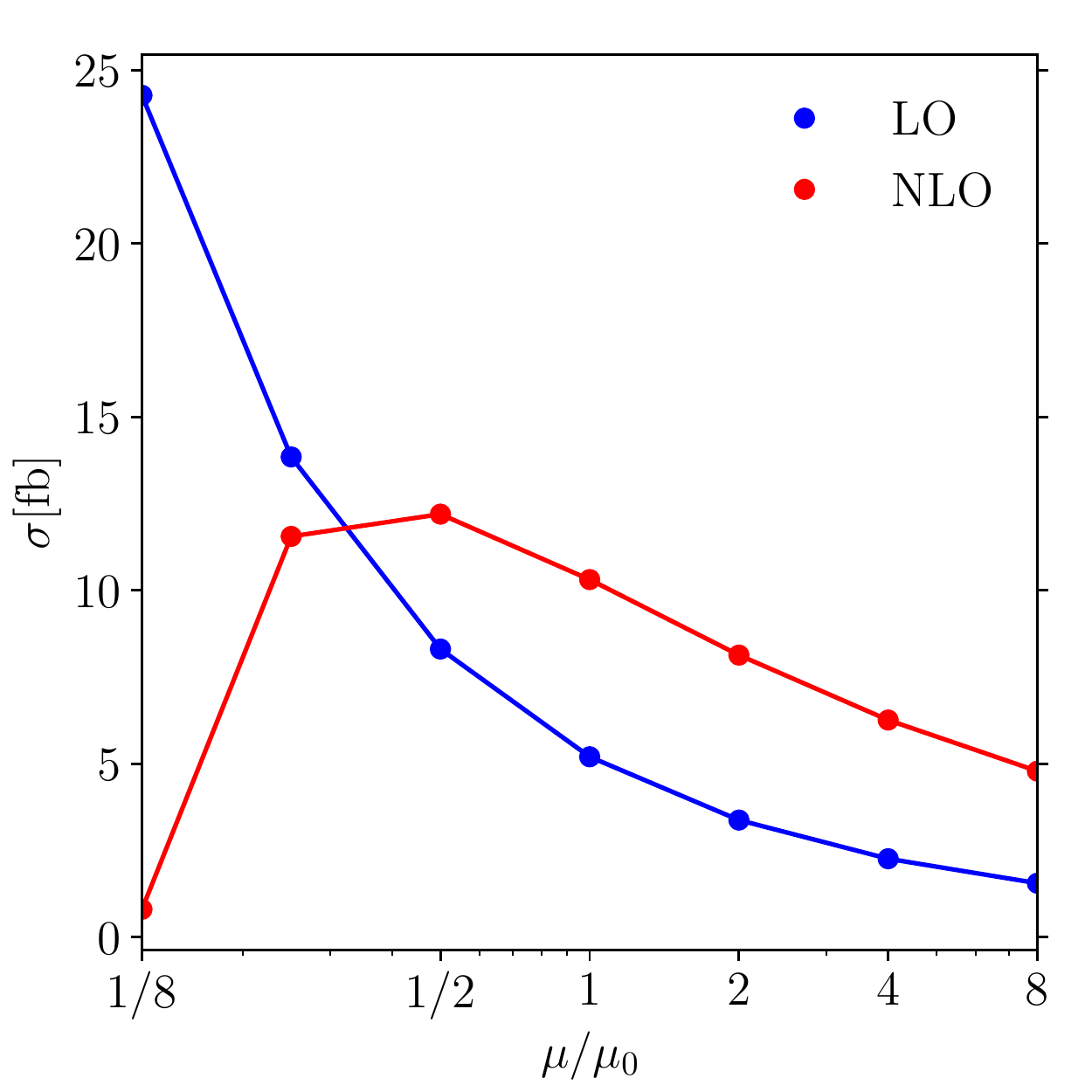}
        \vspace*{-1ex}
        \caption{\label{fig:xs}
          Cross section at LO and NLO in fb for the process $\Pp\Pp
          \to \mu^-\bar\nu_{\mu}\Pe^+\nu_{\Pe}\bar\Pb\Pb\bar\Pb\Pb$ at
          $\sqrt{s}=13\TeV$ as a function of the scale $\mu$,
          which refers to both the factorisation and renormalisation scales.
          The central scale $\mu_0$ is defined in Eq.~\refeq{eq:scale}.}
\end{figure}

The LO and NLO cross sections obtained from the full computation read
\begin{equation}\label{eq:sifull}
 \sigma_{\rm LO} = 5.203(4)^{+60\%}_{-35\%} \fb \qquad {\rm and} \qquad \sigma_{\rm NLO} = 10.31(8)^{+18\%}_{-21\%} \fb,
\end{equation}
respectively.  The digits in parentheses indicate the numerical Monte
Carlo errors on the predictions.  The superscript and subscript
represent the percentage scale variations.  We use the conventional
7-point scale variation, \ie we calculate the quantities for the
following pairs of renormalisation and factorisation scales,
\begin{equation}
(\mu_\mathrm{ren}/\mu_0,\mu_\mathrm{fact}/\mu_0) = (0.5,0.5),
(0.5,1),(1,0.5),(1,1),(1,2),(2,1),(2,2)
\end{equation}
with the central scale defined in Eq.~\refeq{eq:scale} and use the
resulting envelope.  The first observation is that the corrections are
substantial and amount to $97.8\%$ for the central scale, \ie the
$K$-factor is $1.98$.  The large $K$-factor is related to our scale choice
\refeq{eq:scale} which results in somewhat larger scales than usual.%
\footnote{We note that the LO cross section scales with
    $\alphas^4$, which results not only in a scale uncertainty of the
    order of 50\% but also in a large variation of $K$-factors. In
    fact, a $K$-factor near 2 is not unusual for this process
    \cite{Jezo:2018yaf,Bevilacqua:2021cit}, if the PDFs used for the
    LO calculation employ the same values for $\alphas$ as those for
    the NLO calculation.}  The dependence of the results on the scale choice is shown
in \reffi{fig:xs} for the case when both the renormalisation and
factorisation scales are set to a common value.
Based on these results, choosing $\mu_0/2$ as central scale might be
preferable, as it is closer to the maximum of the NLO curve and 
gives rise to smaller NLO QCD corrections ($K=1.47$) in agreement with results for
on-shell top quarks in the literature
\cite{Bredenstein:2010rs,Cascioli:2013gfa,Jezo:2018yaf,Buccioni:2019plc}.
In any case, the inclusion of NLO QCD corrections significantly
reduces the size of the scale uncertainty from $[+60\%, -35\%]$ to
$[+18\%, -21\%]$.

Table \ref{table:crossection} shows the cross sections of the
different partonic channels.
\begin{table}
\begin{center}
\begin{tabular}{ c  c  c  c  c }
 Ch. & $\sigma_{\rm LO}$ [fb] & $\sigma_{\rm NLO}$ [fb] & $K$-factor & $\delta [\%]$\\
  \hline\hline
 $\Pg\Pg$              & $4.861(4)$   & $9.95(8)$   & $2.05$ & $96.4$\\
 $\Pq\bar{\Pq}$        & $0.3298(1)$  & $0.43(1)$   & $1.30$ & $4.2$ \\
   $\Pb \bar{\Pb}$     & $0.00742(1)$ & $0.017(2)$  & $2.29$ & $0.16$ \\
$\Pg\Pq/\Pg\bar{\Pq}$  & -            & $-0.19(2)~$ & -      & $-1.8~$ \\
$\Pg\Pb/\Pg\bar{\Pb}$  & -            & $0.094(2)$  & -      & $0.91$ \\
$\Pb\Pb$               & $0.00263(1)$ & $0.0072(9)$  & $2.76$ & $0.070$ \\
$\bar\Pb\bar\Pb$       & $0.00262(1)$ & $0.0057(8)$  & $2.18$ & $0.055$ \\
  \hline
   $\Pp\Pp$            & $5.203(4)$   & $10.31(8)$  & $1.98$ & $100$ \\
  \hline
\end{tabular}
\end{center}
\caption{
Fiducial cross sections at LO and NLO in fb for the process $\Pp\Pp \to \mu^-\bar\nu_{\mu}\Pe^+\nu_{\Pe}\bar\Pb\Pb\bar\Pb\Pb$ with its corresponding sub-channels at $\sqrt{s}=13\TeV$.
The channels for light quark flavours $\Pq=\Pu,\Pd,\Pc,\Ps$ are grouped into one category.
The quark--gluon channels denoted by $\Pg\Pq/\Pg\bar{\Pq}$ appear only at NLO in the real corrections.
The contributions involving bottom quarks in the initial state, $\Pb
\bar{\Pb}$, $\Pg\Pb/\Pg\bar{\Pb}$, $\Pb\Pb$, and  $\bar\Pb\bar{\Pb}$, are shown separately.
The hadronic cross section is listed in the last line of the table
dubbed $\Pp\Pp$. The $K$-factors are defined as $K = \sigma_{\rm NLO} / \sigma_{\rm LO}$, and $\delta$ represents the contributions relative to the full NLO result.
The integration errors of the last digits are given in parentheses.}
\label{table:crossection}
\end{table}
As usual at the LHC, the gluon-initiated
contributions largely dominate the partonic cross section.  For
example, at LO the $\Pg\Pg$ channel represents $96.4\%$ of the hardonic
cross section, while the $\Pq \bar{\Pq}$ channels with $
q=\Pu,\Pd,\Pc,\Ps$ give $4.2\%$ and $\Pb \bar{\Pb}$, $\Pb\Pb$,
and $\bar\Pb \bar{\Pb}$, only
$0.16\%$, $0.070\%$, and $0.055\%$, respectively.
The $\Pg\Pg$ and $\Pq \bar{\Pq}$ channels get NLO QCD
$K$-factors $2.05$ and $1.30$, respectively.  Such differences have
already been observed for several top--antitop production processes
(see for instance \citeres{Bredenstein:2009aj,Denner:2015yca,Denner:2017kzu}).
We note that for the bottom-induced channels, the $K$-factor is even
higher and ranges between $2.18$--$2.76$. However, these contributions are
greatly suppressed by their (anti-)bottom-quark
PDFs and are below $0.3\%$ of the total cross section at both LO and NLO.
At NLO, new partonic channels are opening up.  The $\Pg\Pq/\Pg\bar{\Pq}$ channels
yield rather small negative corrections (of the order of $-1.8\%$ of
the total NLO cross section), while the $\Pg\Pb/\Pg\bar{\Pb}$
contribution is positive and reaches $+0.9\%$.
Overall the NLO corrections are dominated by the ones of the $\Pg \Pg$ channel
to raise a $K$-factor of $1.98$. 

The cross sections in the tt-DPA, retaining only doubly-top-resonant
contributions {as specified in \refse{se:DPA}}, read
\begin{equation}
\sigma_{\rm LO}^{\rm DPA} = 5.029(2)^{+60\%}_{-35\%} \fb \qquad {\rm and}
\qquad \sigma_{\rm NLO}^{\rm DPA} = 10.23(8)^{+19\%}_{-21\%} \fb.
\end{equation}
These values should be compared to the ones of the full computation in Eq.~\refeq{eq:sifull}.
First, the scale variation is essentially the same as in the full
calculation, indicating that the functional dependence of the cross
sections on the renormalisation and factorisation scales is not
significantly modified.  Looking at the central values, one observes
that the tt-DPA is lower than the full computation by $3.3\%$ at LO.
At NLO, the difference between the two cross sections ($0.8\%$) is of
the order of the integration error which is $0.7\%$.  This is due to
the way the NLO DPA is constructed (see
\refse{se:description_computation}).  While the full LO and real
contributions are used, only the virtual contributions are computed in
the pole approximation.  Since the virtual corrections amount to about
$30\%$, the expected error of the tt-DPA at NLO is about $30\%$ of the
expected error at LO, \ie $0.3\Gt/\Mt\approx 0.25\%$.

\begin{figure}
        \begin{subfigure}{0.49\textwidth}
                \includegraphics[width=\textwidth]{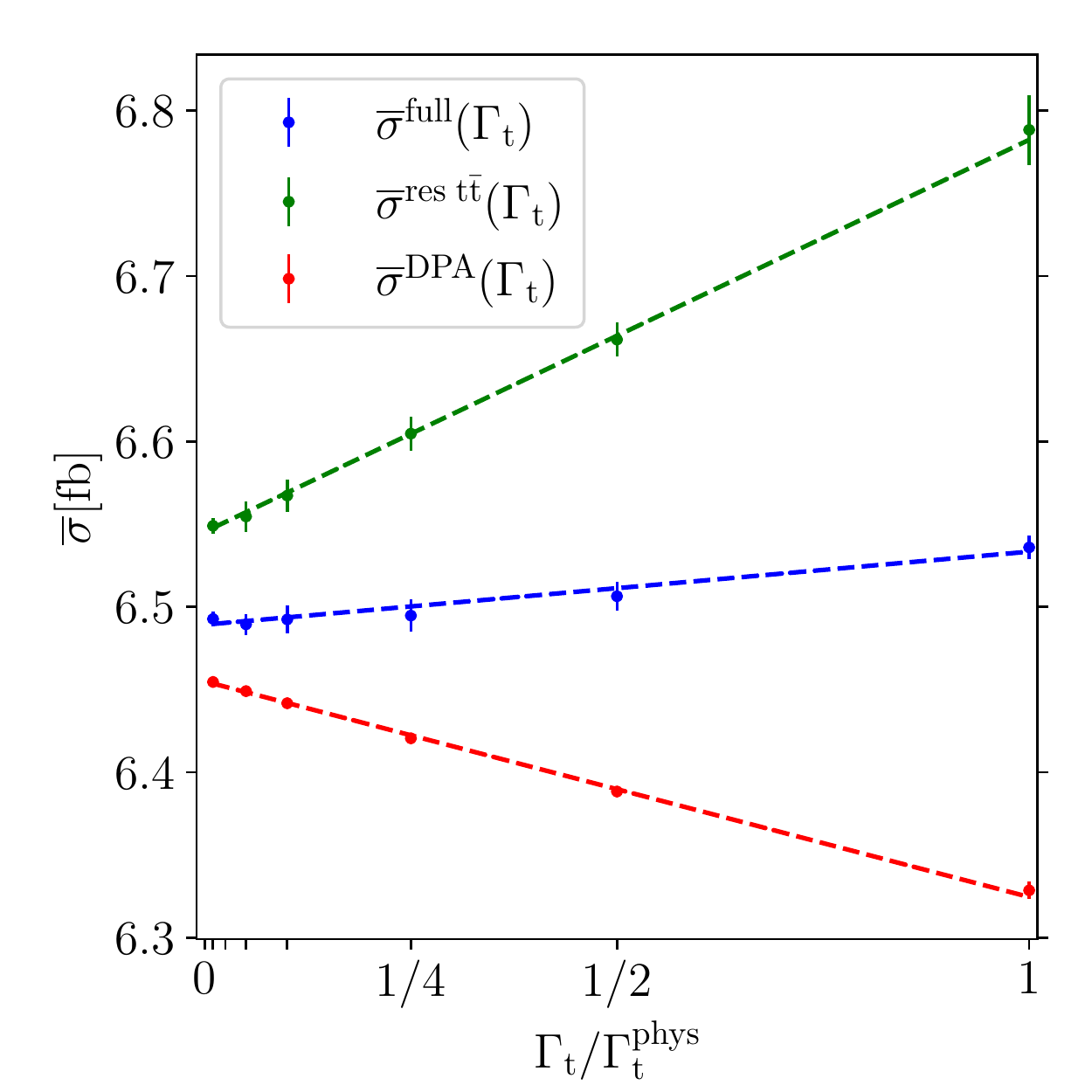}
        \end{subfigure}
        \hfill
        \begin{subfigure}{0.49\textwidth}
                \includegraphics[width=\textwidth]{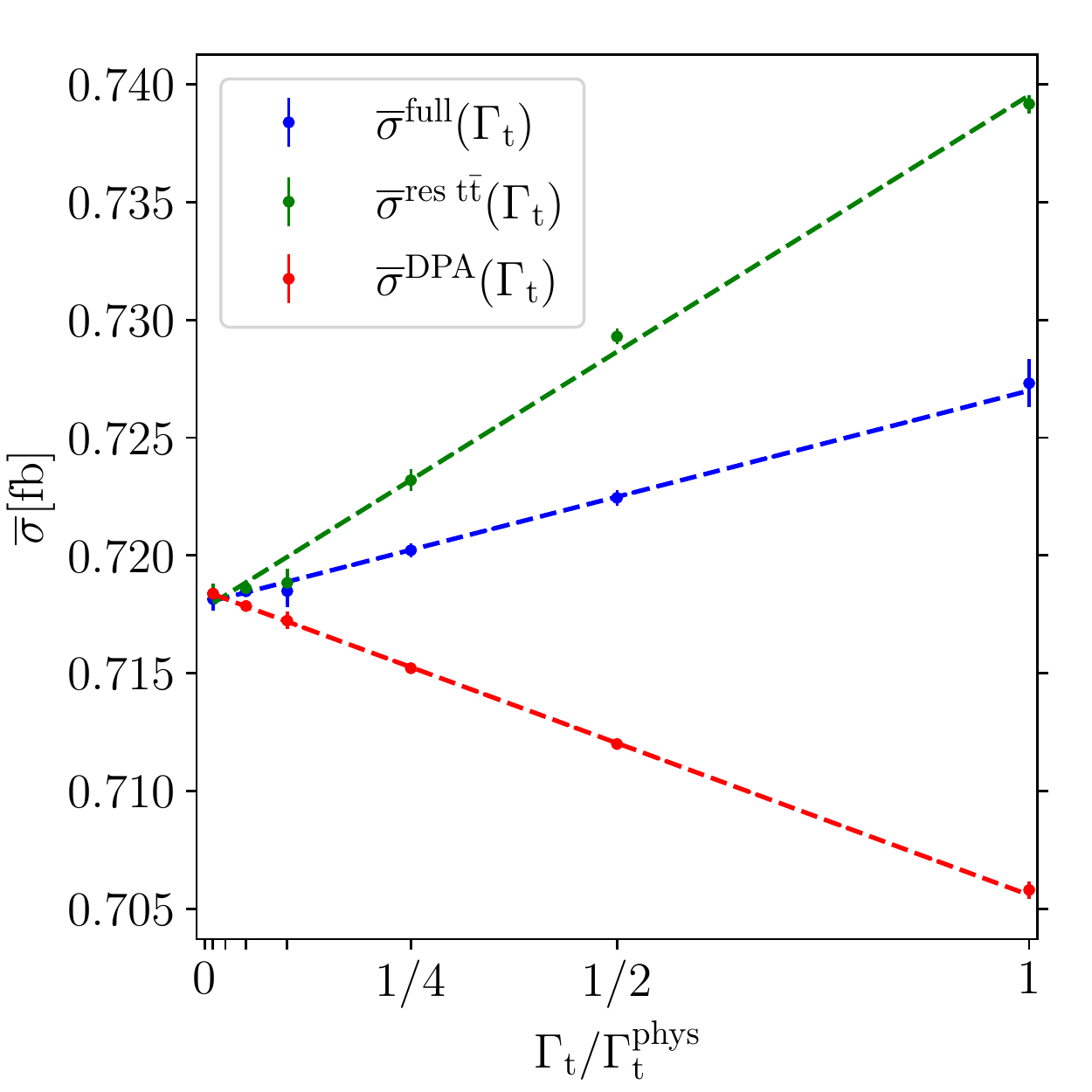}
        \end{subfigure}

        \caption{
          \label{fig:nwa}.
          Extrapolation of fiducial cross sections to the limit $\Gamma_\Pt\to0$
          for the gluon-induced channel 
          $\Pg\Pg\to\mu^+\nu_\mu\Pe^+\nu_{\Pe}\bar\Pb\Pb\bar\Pb\Pb$
          for a fixed scale $\mu_{\mathrm{ren}}=\mu_{\mathrm{fact}}= 173.34\GeV$.
          The left plot shows results for our default setup (but fixed
          scale) while in 
          the plot on the right we imposed an additional invariant-mass
          cut $M_{\Pb\Pb}>110\GeV$.           }
\end{figure}

For $\Pt\bar\Pt$ production, finite-width effects at the level of one
percent have been found by comparing the off-shell calculation with
the narrow-top-width limit in
\citeres{Bevilacqua:2010qb,Denner:2012yc,AlcarazMaestre:2012vp}.  It
is instructive to perform a similar analysis for the process
\refeq{eq:LOprocess}. To this end, we determine the corresponding
narrow-top-width limit in LO as in \citere{Denner:2012yc} by a numerical
extrapolation,
$$
\bar\sigma(\Gamma_\Pt) = \sigma(\Gamma_\Pt)
\left(\frac{\Gamma_\Pt}{\Gamma_\Pt^{\mathrm{phys}}}\right)^2
$$
in the range $0<\Gamma_\Pt<\Gamma_\Pt^{\mathrm{phys}}$, where
$\Gamma_\Pt^{\mathrm{phys}}$ is the physical top-quark width from
Eq.~\refeq{eq:mt}, for the dominating $\Pg\Pg$ channel. The factor
$\left(\Gamma_\Pt/\Gamma_\Pt^{\mathrm{phys}}\right)^2$ ensures that
effective top-decay branching ratios remain constant in the limit. In
\reffi{fig:nwa} we show results of this extrapolation for the full
calculation, for the DPA, and for a calculation (res $\Pt\bar\Pt$)
where we take into account the same subset of diagrams as in the DPA
but do not perform the on-shell projection.
This analysis yields two interesting results: First, the difference
between the DPA and the full calculation is larger than the difference
between the full calculation and its narrow-width approximation. This
indicates that contributions of non-resonant diagrams and finite-width
effects of the resonant diagrams cancel to some extent, and the
generic accuracy of on-shell calculations is worse. Second, the
narrow-width limits of the three calculations do not agree, but differ
at the level of one percent, the size of finite-width effects. The
origin of these differences are contributions of resonant
top or antitop quarks that decay via $\Pt\to\nu_\Pe\Pe^+\Pb\bar\Pb\Pb$
or $\bar\Pt\to\bar\nu_\mu\mu^-\bar\Pb\bar\Pb\Pb$. While some of these
contributions, \eg \reffi{fig:diag_ttg}, are included in the subset
of diagrams selected for the DPA, others, \eg \reffi{fig:diag_twbg},
are not. This demonstrates that the narrow-width limit of the full
calculation differs by contributions of the order of $\Gamma_\Pt/\Mt$
from an on-shell calculation of $\Pt\bar\Pt\Pb\bar\Pb$ production with
subsequent top decays. The construction of a narrow-width
approximation based on on-shell calculations that takes into account
all these top resonances would be non-trivial. While the
  extrapolation was performed at LO, the same kind of features
  persists in an NLO calculation.

The situation is different for $\Pt\bar\Pt$ production, where no extra
top resonances are present. We verified with our codes that for
$\Pg\Pg\to\mu^-\bar\nu_{\mu}\Pe^+\nu_{\Pe}\bar\Pb\Pb$ the DPA and the
approximation ``res $\Pt\bar\Pt$'' based on resonant $\Pt\bar\Pt$
diagrams approach the same value in the narrow-top-width limit.  Also
for $\Pt\bar\Pt\Pb\bar\Pb$ production the cross sections for the
different approximations in the narrow-top-width limit coincide, if
one eliminates the resonances related to
$\Pt\to\nu_\Pe\Pe^+\Pb\bar\Pb\Pb$ or
$\bar\Pt\to\bar\nu_\mu\mu^-\bar\Pb\bar\Pb\Pb$ decays with additional
cuts. This is demonstrated in the right plot in \reffi{fig:nwa}, where we
imposed the additional cut $M_{\Pb\Pb}>110\GeV$ on all pairs of bottom
and/or antibottom jets.

\subsection{Differential distributions}

\begin{figure}
        \begin{subfigure}{0.49\textwidth}
                \includegraphics[width=\textwidth]{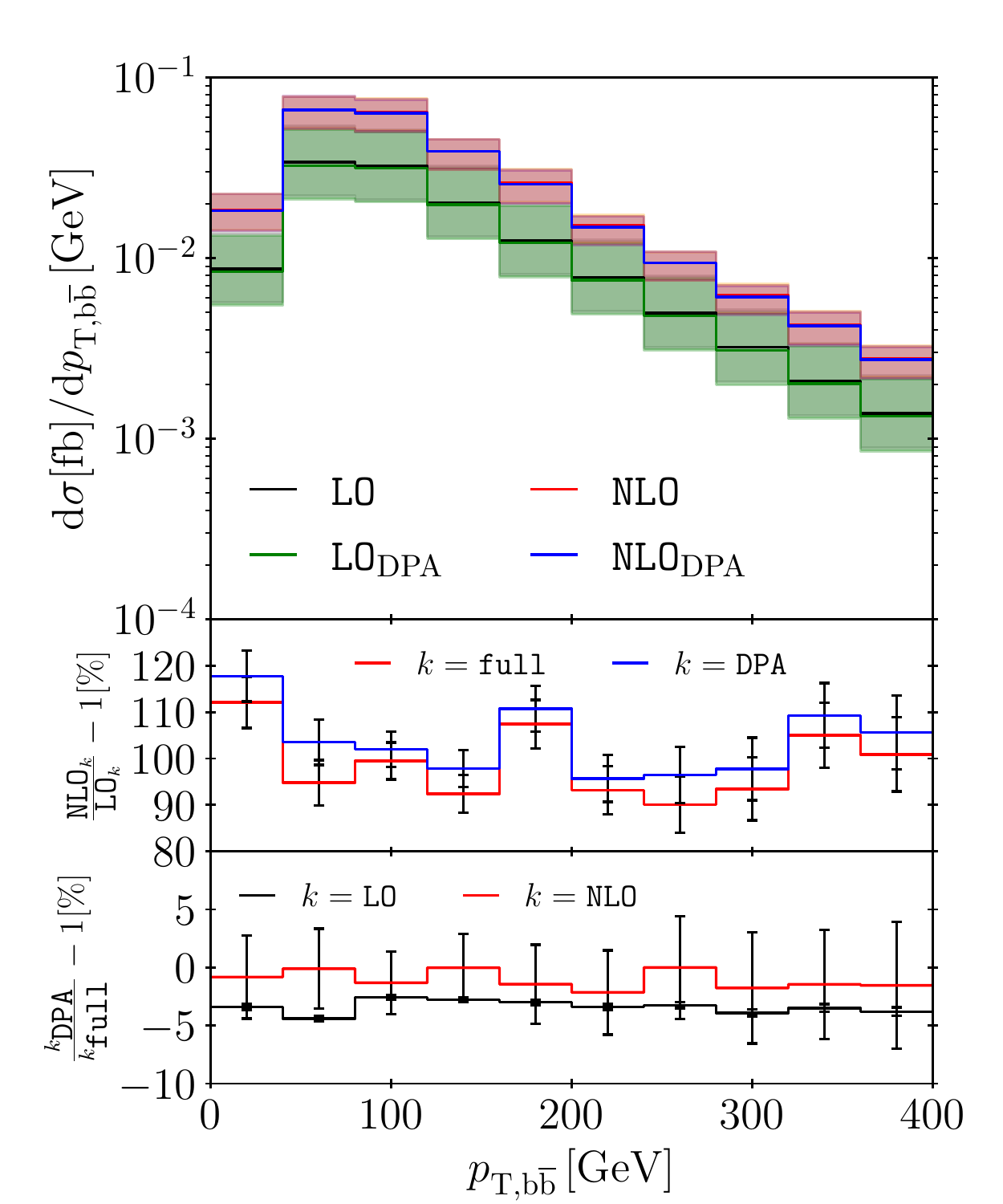}
        \end{subfigure}
        \hfill
        \begin{subfigure}{0.49\textwidth}
                \includegraphics[width=\textwidth]{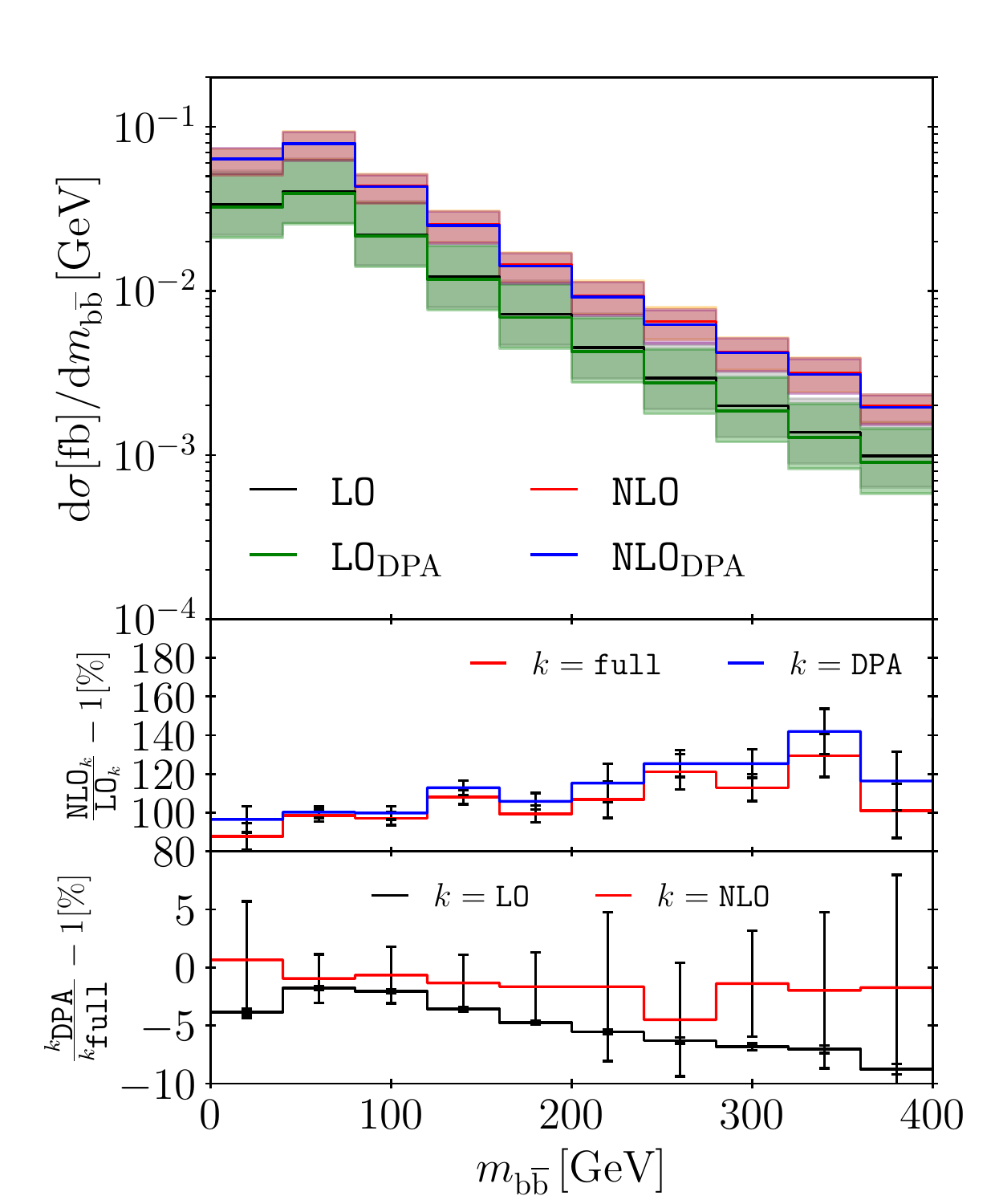}
        \end{subfigure}

        \caption{\label{fig:dist_bb}
          Differential distributions at LO and NLO for $\Pp\Pp\to\mu^+\nu_\mu\Pe^+\nu_{\Pe}\bar\Pb\Pb\bar\Pb\Pb$:
          transverse momentum of the two bottom jets not originating from a top quark,
          and invariant mass of the two bottom jets not originating from a top quark.}
\end{figure}

\begin{figure}
        \begin{subfigure}{0.49\textwidth}
                \includegraphics[width=0.9\textwidth]{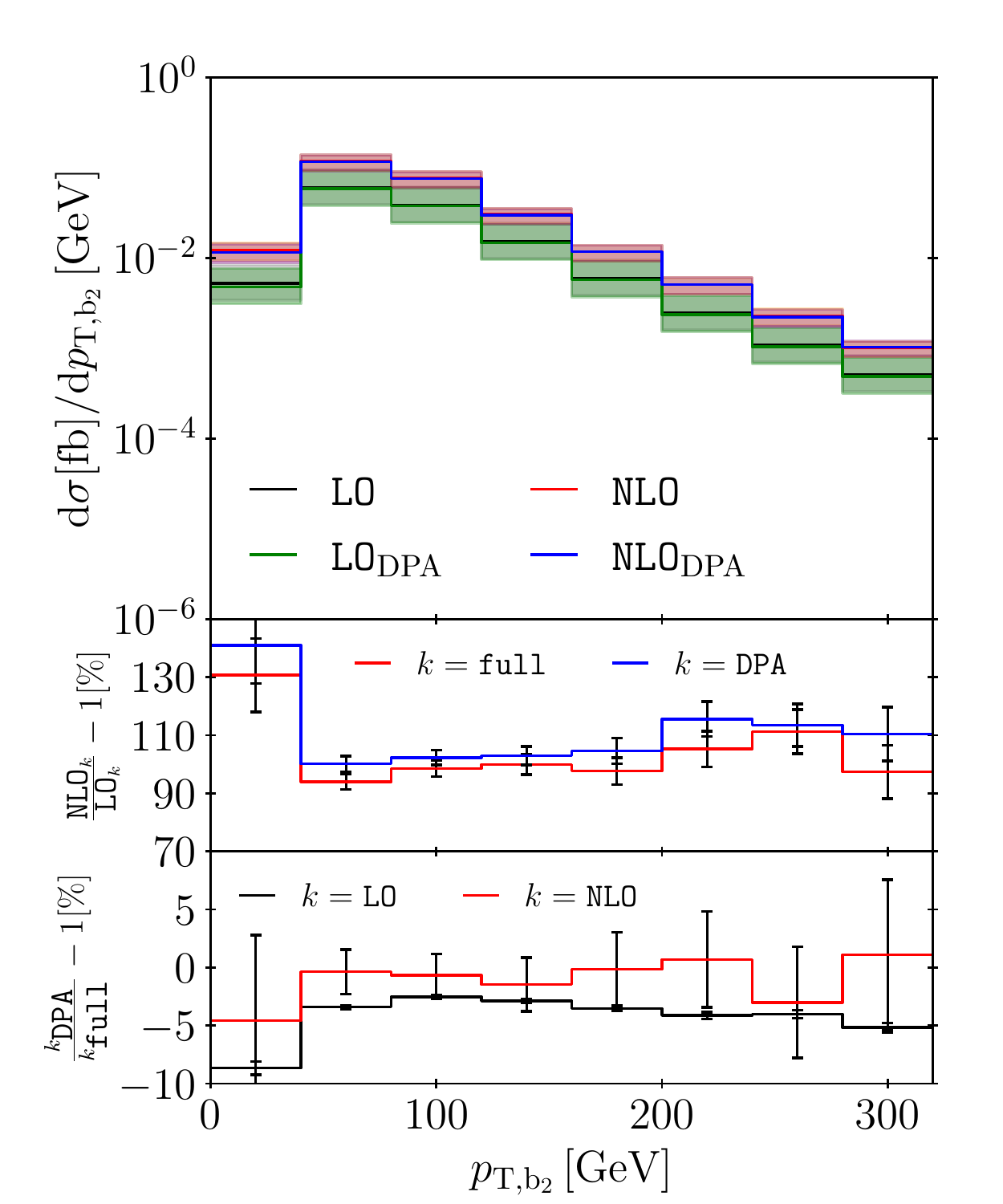}
        \end{subfigure}
        \hfill
        \begin{subfigure}{0.49\textwidth}
                \includegraphics[width=0.9\textwidth]{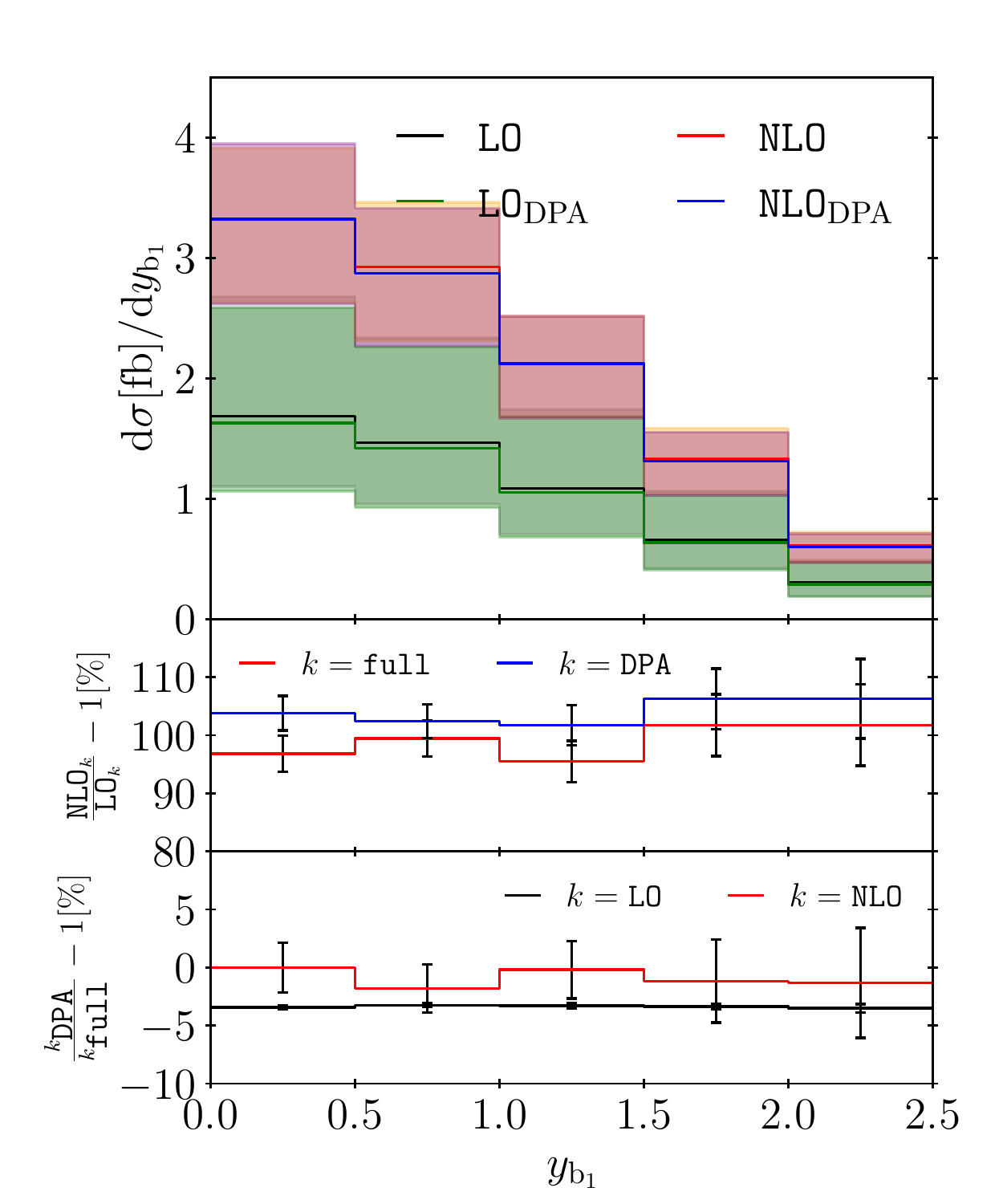}
        \end{subfigure}

        \begin{subfigure}{0.49\textwidth}
                \includegraphics[width=0.9\textwidth]{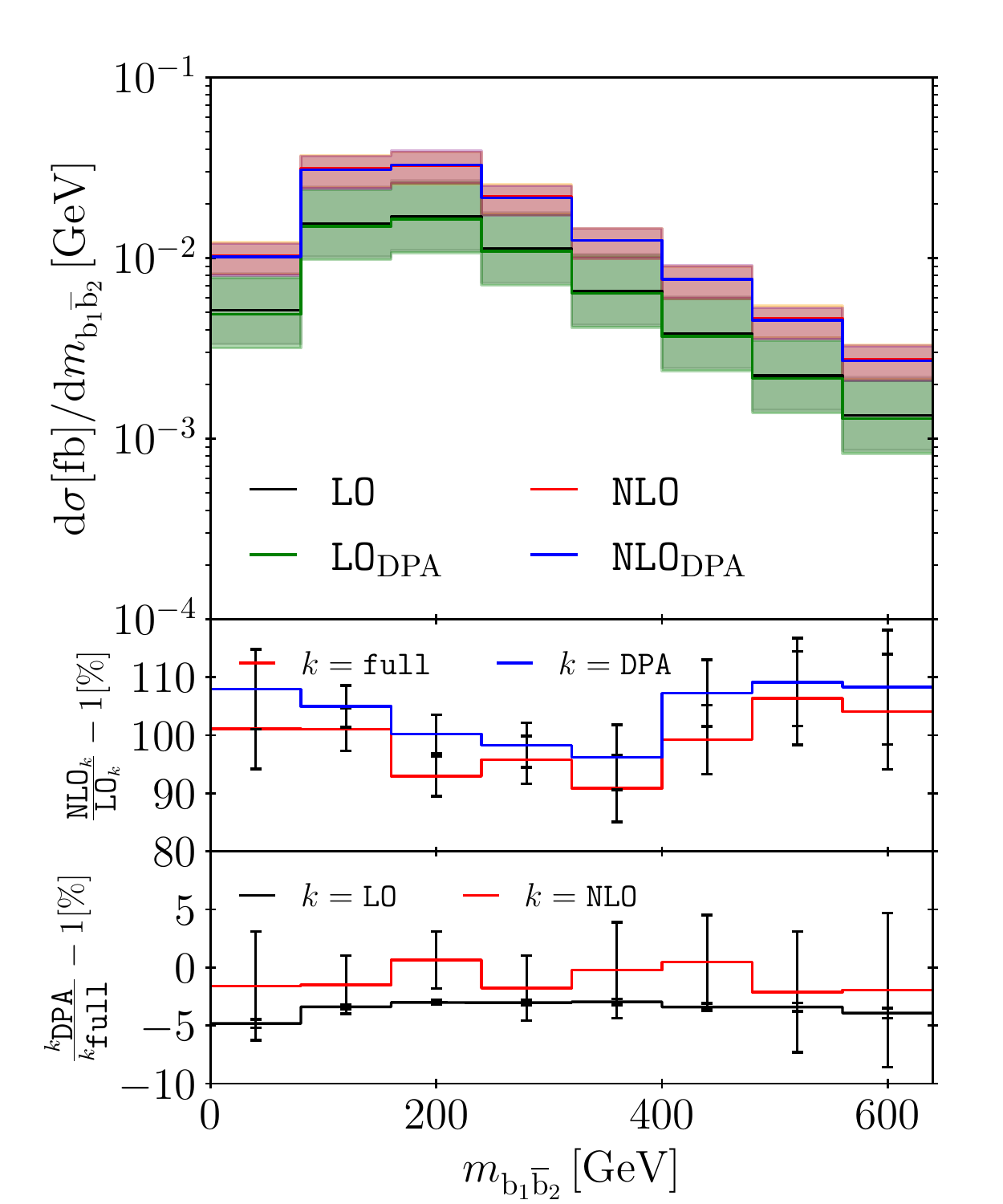}
        \end{subfigure}
        \hfill
        \begin{subfigure}{0.49\textwidth}
                \includegraphics[width=0.9\textwidth]{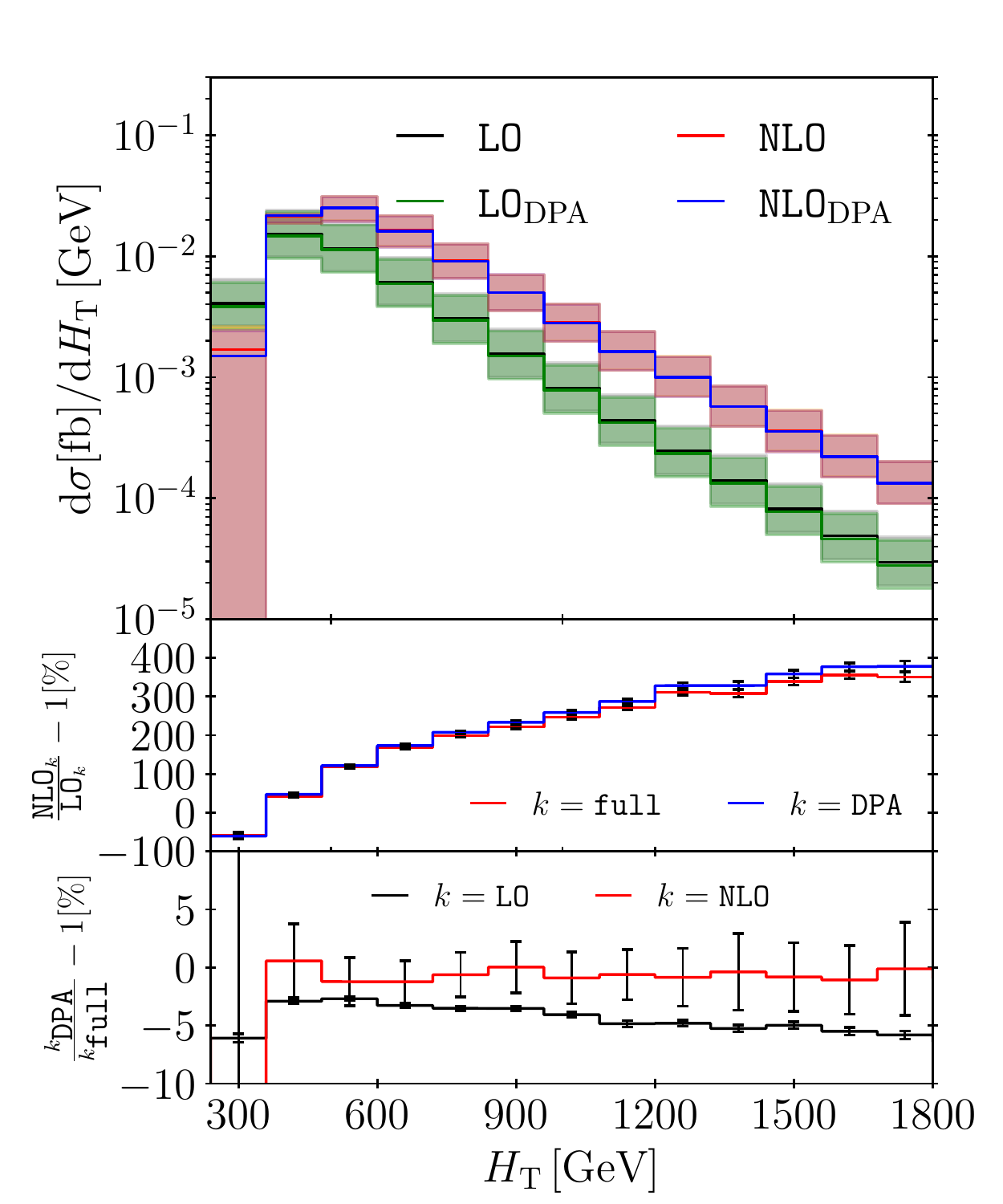}
        \end{subfigure}

        \caption{\label{fig:dist_jet}
          Differential distributions at LO and NLO for $\Pp\Pp\to\mu^+\nu_\mu\Pe^+\nu_{\Pe}\bar\Pb\Pb\bar\Pb\Pb$:
          transverse momentum of the second-hardest b~jet,
          rapidity of the hardest b~jet,
          invariant mass of the hardest and second-hardest b~jet,
          and $H_{\rm T}$ observable (see text for definition).}
\end{figure}

\begin{figure}
        \begin{subfigure}{0.49\textwidth}
                \includegraphics[width=0.9\textwidth]{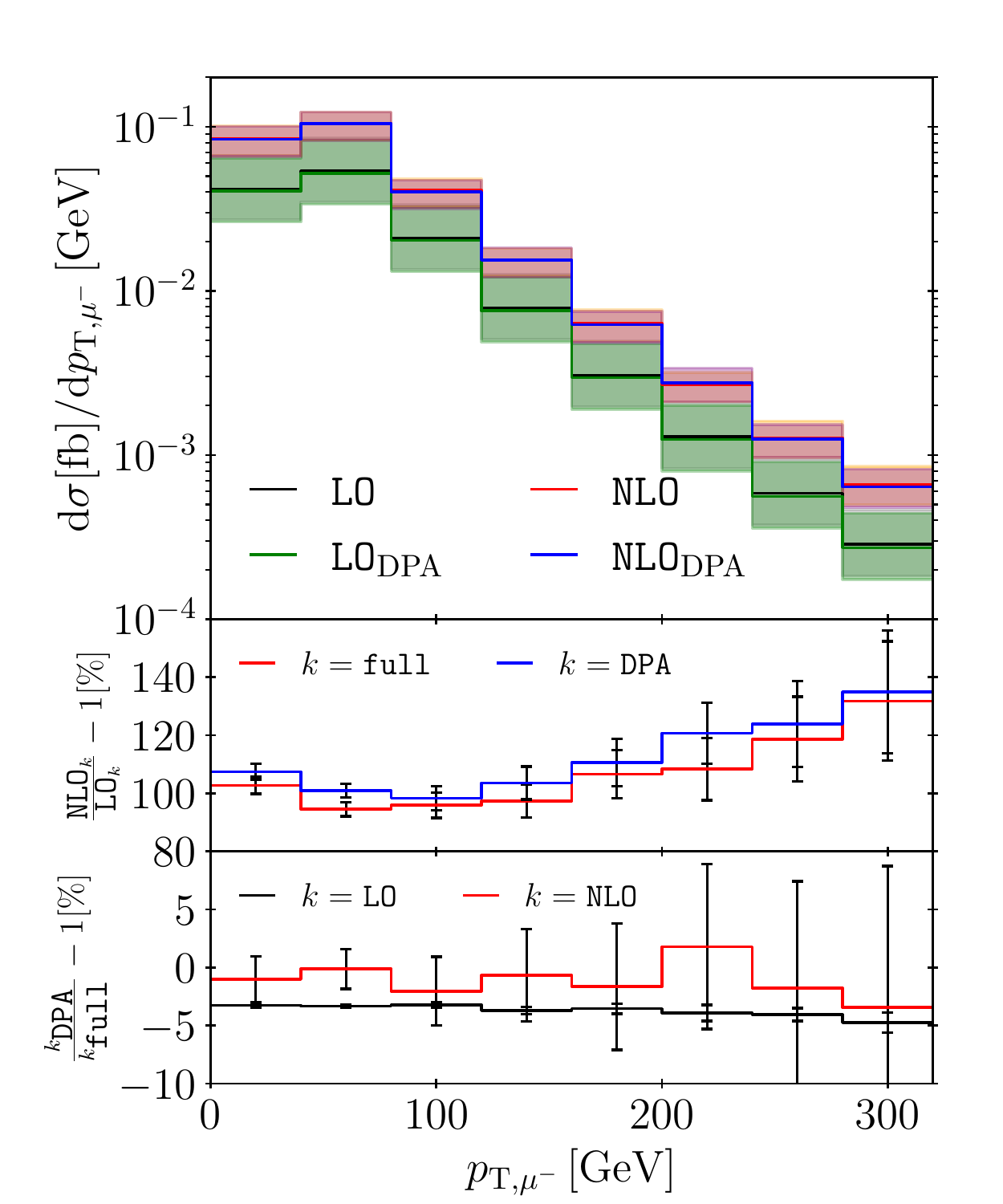}
        \end{subfigure}
        \hfill
        \begin{subfigure}{0.49\textwidth}
                \includegraphics[width=0.9\textwidth]{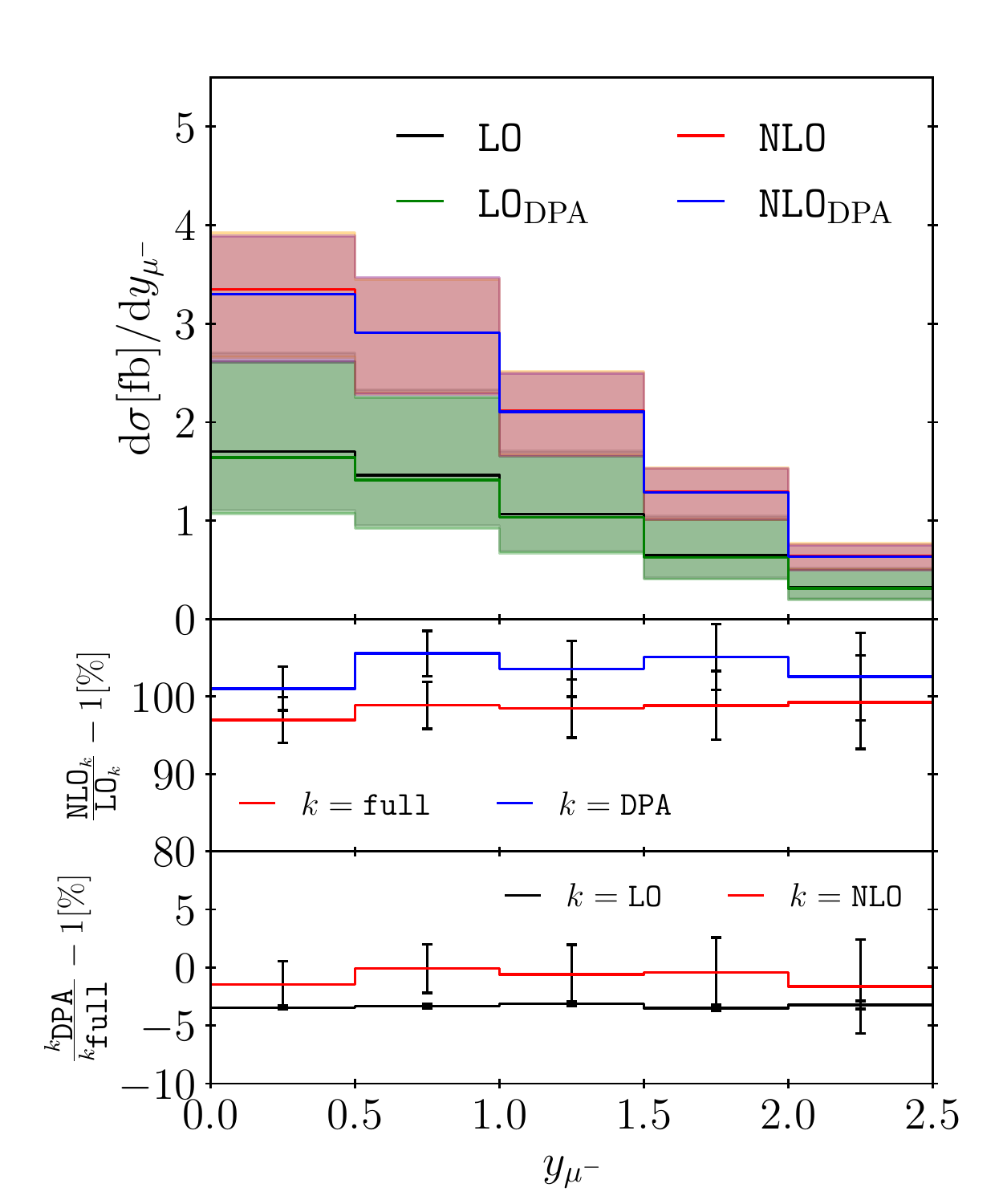}
        \end{subfigure}

        \begin{subfigure}{0.49\textwidth}
                \includegraphics[width=0.9\textwidth]{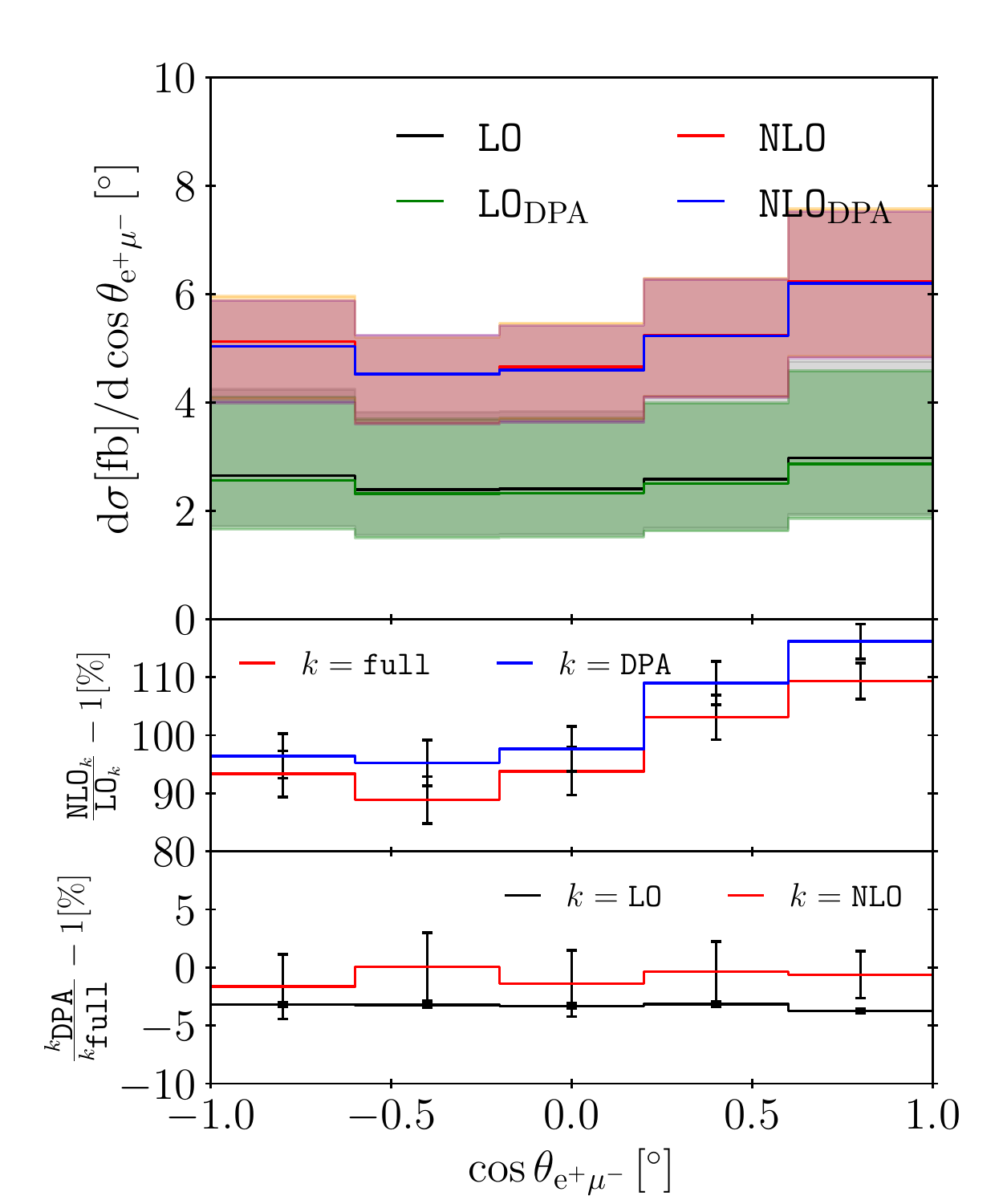}
        \end{subfigure}
        \hfill
        \begin{subfigure}{0.49\textwidth}
                \includegraphics[width=0.9\textwidth]{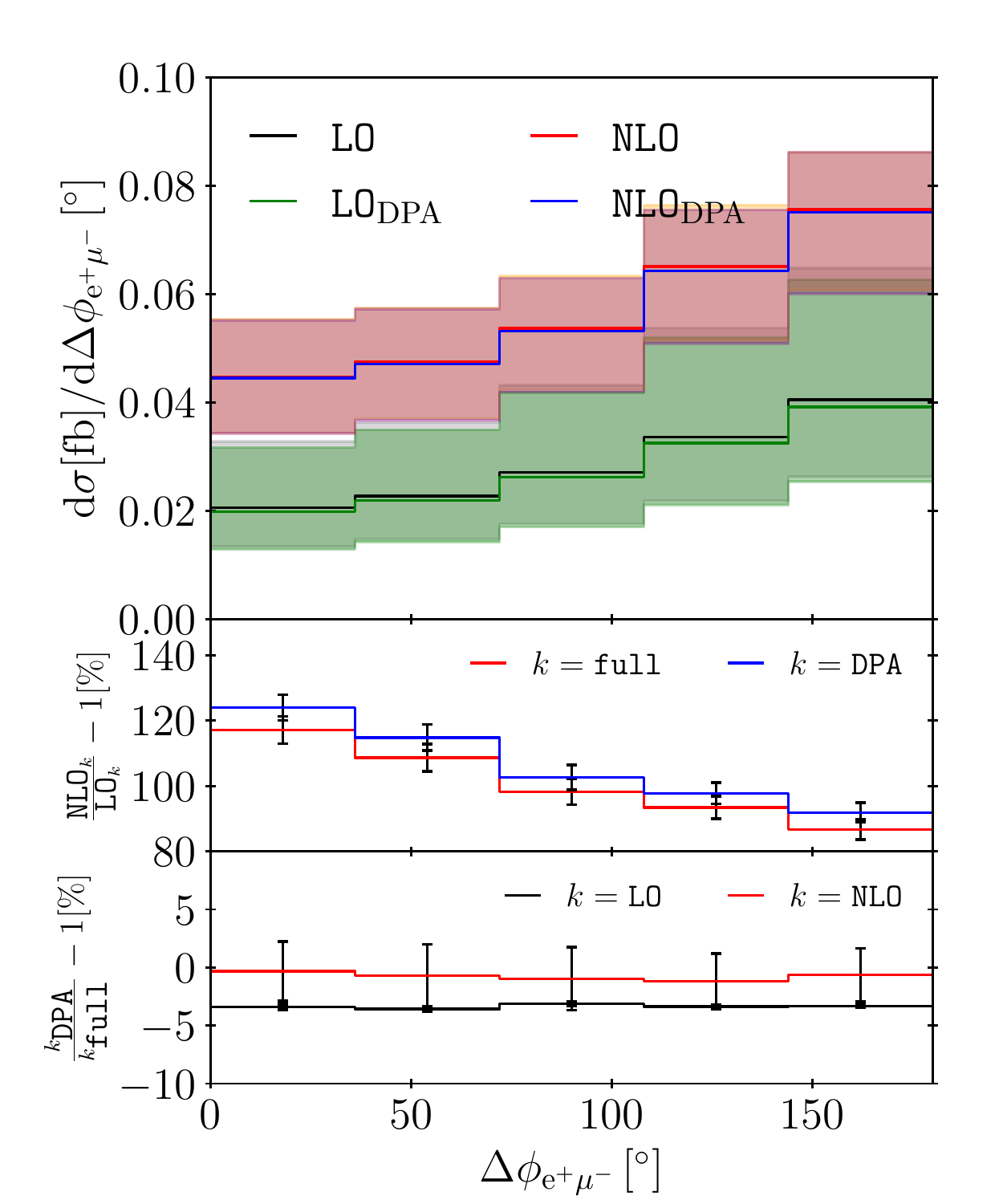}
        \end{subfigure}

        \caption{\label{fig:dist_lep}
          Differential distributions at LO and NLO for $\Pp\Pp\to\mu^+\nu_\mu\Pe^+\nu_{\Pe}\bar\Pb\Pb\bar\Pb\Pb$:
          transverse momentum of the muon,
          rapidity of the muon,
          cosine of the angle between the muon and the positron,
          and azimuthal-angle distance between the muon and the positron.}
\end{figure}

Turning to differential distributions, several physical
observables are shown in \reffis{fig:dist_bb}--\ref{fig:dist_lep}.
While in the upper panels, the absolute predictions at LO and NLO QCD
in the full and in the tt-DPA are displayed, the two lower panels show
the same contributions with respect to different normalisations.
In these panels the error bars represent the numerical Monte Carlo errors.
In the middle insets the size of the QCD corrections in the full
computation and in the tt-DPA is compared.  Finally, the lower insets
illustrate the quality of the approximate computation by normalising the
tt-DPA to the full computation at both LO and NLO QCD.

In \reffi{fig:dist_bb}, we present the  distributions in the
transverse momentum and the invariant mass of the bottom--antibottom
pair that does not originate from the top-quark decay.
In a $\Pt\bar\Pt\PH$ analysis with $\PH\to\Pb\bar\Pb$,
this pair of bottom jets corresponds to the background of the
decay products of the Higgs boson.  More precisely, the bottom jets
are identified as originating from a top quark by maximising the
likelihood function $\mathcal{L}$, defined as a product of two
Breit--Wigner distributions corresponding to the top-quark and
antitop-quark propagators,
\begin{equation}
 \mathcal{L}_{ij} = \frac{1}{\left(p^2_{\mu^-\bar\nu_\mu\Pb_i} -
     m_\Pt^2\right)^2+\left(m_\Pt \Gamma_\Pt\right)^2} \;
 \frac{1}{\left(p^2_{\Pe^+\nu_\Pe\Pb_j} - m_\Pt^2\right)^2+\left(m_\Pt \Gamma_\Pt\right)^2} ,
\end{equation}
where the momenta $p_{a b c}$ are defined as $p_{a b c} = p_{a} +
p_{b} + p_{c}$.  The combination of bottom jets $\{\Pb_i, \Pb_j \}$
that maximises this function defines the two bottom jets originating
from top quarks.  From the 2 or 3 bottom jets left in the event, the
two hardest ones, \ie those with highest  transverse momenta,
define the bottom--antibottom pair that does not originate
from the top-quark decay and whose transverse-momentum and
invariant-mass distributions are shown in \reffi{fig:dist_bb}.
The distribution in the transverse momentum of the two bottom jets
not coming from a top decay shows rather stable corrections around
$100\%$ apart from low transverse momentum, where the QCD corrections
reach $110\%$.  The difference between the full calculation and the
one in DPA does not show significant variations over the phase space
neither at LO nor at NLO QCD but is largely inherited from the
fiducial cross section. In particular, the difference between the tt-DPA
and the full calculation at NLO is within the integration errors, as
for all following distributions.
The distribution in the invariant mass of the bottom--antibottom pair,
on the other hand, exhibits larger variations between the full
computation and the tt-DPA one at LO.  The difference between the two
computations is about $4\%$ in the first bin, decreases to a few per
cent around $100\GeV$ where the bulk of the cross section is located,
and increases to almost $10\%$ at $400\GeV$.  
The additional contributions not contained in the tt-DPA increase the
cross section. The largest effects appear for small
$m_{\Pb\bar\Pb}$, a region that is enhanced by $\Pb\bar\Pb$ pairs
resulting from virtual gluons (for instance \reffi{fig:diag_1resbb}),%
\footnote{Note, however, that the jet-resolution parameter $R=0.4$
  together with the transverse-momentum cut on the b~jets of $25\GeV$
  imply a minimum invariant mass of two b~jets of about $10\GeV$.}
and for large invariant masses, where
diagrams with bottom quarks coupling directly to the incoming gluons
give sizeable contributions (for instance \reffi{fig:diag_1restb}).
 The QCD corrections tend to grow when going to
higher invariant masses.  This is in contrast to the results of
on-shell computations (see Figures 6 and 17 of
\citere{Bredenstein:2010rs}), where the relative corrections
to the invariant-mass distribution tend to decrease with
increasing invariant mass. This is most likely due to the different
scale choices in the on-shell and off-shell calculations, where the
scales in the latter tend to be higher.
We mention that the two distributions in
\reffi{fig:dist_bb} are the only ones that can be qualitatively
compared with results of the literature where stable top quarks are
used \cite{Bredenstein:2010rs,Cascioli:2013era,Jezo:2018yaf}.  Since
these computations are, however, done with different event selections
and scale choices,
a direct comparison is rather difficult.  The transverse-momentum
distribution is given as well in Figure 10 of
\citere{Bredenstein:2010rs} but with a cut of $100\GeV$ on the
invariant mass of the two bottom jets. Nevertheless, the differential $K$-factor
is flat for this distribution above $50\GeV$ both in the off-shell and
on-shell calculation.

In \reffi{fig:dist_jet}, the distribution in the transverse momentum
of the second-hardest b~jet is shown.
The full corrections are large (about $130\%$) at low transverse
momentum, then become smaller to finally reach roughly $100\%$ at
$300\GeV$. Such a behaviour has already been observed in top--antitop
production in the lepton+jets channel for the transverse momentum of
the hardest b~jet \cite{Denner:2017kzu}.  The large corrections were
attributed to real radiations that take away momentum of the
final-state particles.  The effect of non-doubly-resonant top quarks
is rather clear in this distribution at LO.  The tt-DPA deviates from
the full computation by almost $10\%$ in the first bin.  The
difference is minimal at $100\GeV$ but always between $2\%$ and $5\%$.
This indicates significant non-doubly-resonant contributions that
might originate from multi-peripheral diagrams where the bottom quarks
couple directly to the incoming gluons (\reffi{fig:diag_1restb}).
Moreover, while bottom jets resulting from top quark decays tend to
have transverse momenta of the order of the top mass, this is not the
case for bottom jets in general.
Looking at the distributions in the transverse momentum of the
  other b~jets (not shown here), the difference between the tt-DPA and
  the full computation is reduced at low transverse momenta of the
  third and fourth hardest b~jet, but enhanced for the hardest one.
  In contrast to the case of the hardest and second hardest b~jets,
  for the third and fourth hardest b~jets these configurations receive
  also contributions with doubly-resonant top quarks that are included
  in the tt-DPA.  At NLO, the differences are within integration
errors owing to the fact that the DPA is only applied to the virtual
amplitudes.
For the distribution in the rapidity of the hardest b~jet, the full
and the approximate computation have the same qualitative behaviour.
The full NLO QCD corrections are essentially flat in this distribution.
They are a bit above $+100\%$ at rapidity $2.5$ and slightly below $+100\%$ in
the central region.
The distribution in the invariant mass of the two hardest bottom jets is depicted in the bottom left of \reffi{fig:dist_jet}.  These
bottom jets can either originate from a top-quark decay or are produced
directly.  The corrections tend to be larger at low and large invariant
mass ($100\%$ at $0\GeV$ and $105\%$ at $600\GeV$)
and reach a minimum around $300\GeV$ of $95\%$.
The quality of the tt-DPA is rather good in this observable in the sense that
no significant shape distortion is observed and only a difference in
the overall normalisation is present.
The last plot in \reffi{fig:dist_jet} concerns the distribution in
$H_{\rm T}$, defined as
\begin{equation}
 H_{\rm T} = p^{\mathrm{miss}}_{\rT} + \sum_{i=\Pe^+,\mu^-,\Pj_\Pb} E^{i}_{\rT}.
\end{equation}
This observable is interesting because it gives an estimate of the
typical scale of the process.  This is the reason why it enters the
definition of the renormalisation and factorisation scale in
Eq.~\eqref{eq:scale}. Note that as opposed to Eq.~\eqref{eq:scale}, 
here only the four hardest bottom jets fulfilling the event selection
in Eq.~\eqref{eq:cut} are taken into account.
While the corrections are of the usual size for low
values of this quantity, they steadily increase to exceed $300\%$
above $1200\GeV$.  {Such an effect has already been observed for
top-pair production \cite{Denner:2012yc}.}
The quality of the tt-DPA is at the level of $-3\%$ at
LO around the maximum of the distribution near $500\GeV$.  Above and
below, the difference tends to increase: in the first non-trivial bin it amounts
to around $-6\%$, while at high values it steadily reaches $-6\%$ at
$1.8\TeV$.

The final set of distributions shown in \reffi{fig:dist_lep} deals with leptonic observables.
The corrections to the distribution in the transverse momentum of the
muon are larger in the first bin, reach a minimum around $60\GeV$ and 
exceed $130\%$ towards high transverse momentum.  In the same way, the
disagreement between the full computation and the tt-DPA at LO tends to
increase slowly from $3\%$ to $5\%$ when going to large momenta.
Similarly to the distribution in the rapidity of the hardest bottom jet, the one of the muon also does not feature significant shape
distortions in the QCD corrections.  The full corrections are, to a
large extent, inherited from the total cross section. The
  difference between the tt-DPA and the full calculation is flat over
  the kinematic range shown for this observable.
For the distribution in the cosine of the angle between the muon and
positron, the corrections generally tend to increase with
$\cos\theta_{\Pe^+\mu^-}$, ranging between $90\%$ and $110\%$.
The difference between the tt-DPA and the full computations is also flat.
At last, we show the distribution in the azimuthal angle between
the two leptons.  The corrections are at the level of $120\%$ at zero
degree and steadily decrease to $90\%$ in the back-to-back
configuration.  Again, the shape of the tt-DPA computation is quite
similar to the one of the full calculation over the full range.

\section{Conclusion}
\label{se:conclusion}

In this article we have presented the first full NLO QCD computation
for $\Pp\Pp \to\mu^-\bar\nu_{\mu}\Pe^+\nu_{\Pe}\bar\Pb\Pb\bar\Pb\Pb$
at order $\order{\alphas^5 \alpha^4}$ at the LHC.  This final state is
of particular interest as it is shared with
$\Pp\Pp\to\Pt\bar\Pt\PH\bigl(\to\Pb\bar\Pb\bigr)$ which is key for the
extraction of Higgs coupling to top quarks.  The present computation
is carried out with full tree and one-loop matrix elements and, thus,
includes all off-shell and non-resonant contributions.  It therefore
goes beyond the state of the art of fixed-order computations, which
focused so far on the description of the $\Pt\bar\Pt\Pb\bar\Pb$
process with stable top quarks.  In addition to the phenomenological
relevance, the calculation constitutes a significant progress in
complexity as it is the first full NLO QCD computation for a $2\to8$
process with 6 external strongly-interacting particles.  Along with
the full computation, we also provide predictions in a double-pole
approximation which retains only doubly-resonant top contributions in
the virtual corrections.  {Since the Feynman diagrams of the
  considered process include also resonant top and antitop quarks that
  are not related to on-shell production of $\Pt\bar\Pt\Pb\bar\Pb$ and
  subsequent top decays, the narrow-width limit of the full calculation
  differs from an on-shell calculation with subsequent decays, \ie 
  $\Pp\Pp\to\Pt\bar\Pt\Pb\bar\Pb$ followed by
  $\Pt\to\mu^-\bar\nu_{\mu}\Pb$ and $\bar\Pt\to\Pe^+\nu_{\Pe}\bar\Pb$,
by terms of the order of the top width.}
Recent analyses have revealed differences
between various theoretical predictions of $\Pt\bar\Pt\Pb\bar\Pb$ when
including parton-shower effects.  While the present computations
certainly do not lift all discrepancies, they could serve as a basis
for future comparative studies.

At the level of the cross sections, the QCD corrections turn out to be
about $100\%$ for our choice of renormalisation and factorisation
scale.  At the differential level, on top of this overall shift, shape
distortions are present and reach $25\%$ for some distributions.  For
observables that can be compared with on-shell computations
(transverse momentum and invariant mass of the two bottom jets not
coming from the top quarks), we observe qualitative differences in the
shape of the corrections.  This should therefore warrant further
investigations in the future.  At LO, the difference between the full
computation and the one in the double-pole approximation stays below
5\% in most distributions but reaches up to 10\% in some cases.  
Our results show that a simplified calculation using the
  double-pole approximation for the virtual corrections is sufficient
  at this level of accuracy. While it does not provide a quantitative
  statement on the size of off-shell top-quark effects at NLO, it
  nevertheless indicates that these are at least at the level of
  $5$--$10\%$ across phase space.

The results shown here provide an important piece of information
regarding the theoretical description of $\Pt\Pt\Pb\Pb$ production at
hadron colliders.  They should prove useful for present and upcoming
analyses of $\Pp\Pp\to\Pt\bar\Pt\PH\bigl(\to\Pb\bar\Pb\bigr)$ and its
irreducible background at the LHC.

\section*{Note added}
  
  After finishing this work, a similar calculation for the same
  process was published by a different group
  \cite{Bevilacqua:2021cit}. This calculation fully confirmes our
  results and provides an extensive discussion on scale and PDF
  uncertainties. This analysis furthermore revealed that excluding the
  additional jet from the definition of the renormalisation and
  factorisation scale increases the NLO cross section and brings it
  into agreement with the NLO results of other scale definitions
  within the theoretical uncertainties.

\section*{Acknowledgements}

AD acknowledges financial support by the
German Federal Ministry for Education and Research (BMBF) under
contract no.~05H18WWCA1.
JNL was supported by the Swiss National Science Foundation (SNSF)
under contract BSCGI0-157722.
The research of MP has received funding from the European Research Council (ERC) under the European Union's Horizon 2020 Research and Innovation Programme (grant agreement no. 683211).

\bibliographystyle{JHEPmod}
\bibliography{ttbb}

\end{document}

%% file: macros.tex
\def\refeq#1{\mbox{(\ref{#1})}}
\def\reffi#1{\mbox{Figure~\ref{#1}}}
\def\reffis#1{\mbox{Figures~\ref{#1}}}

\def\refse#1{\mbox{Section~\ref{#1}}}

\def\citere#1{\mbox{Ref.~\cite{#1}}}
\def\citeres#1{\mbox{Refs.~\cite{#1}}}

\newcommand{\ie}{\emph{i.e.}\ }
\newcommand{\eg}{\emph{e.g.}\ }

\def\be{\begin{equation}}
\def\ee{\end{equation}}

\newcommand{\Pl}{\ell}
\newcommand{\Pq}{\ensuremath{{q}}}

\newcommand{\PH}{\ensuremath{\mathrm{H}}}
\newcommand{\Pj}{\ensuremath{\mathrm{j}}}
\newcommand{\Pp}{\ensuremath{\mathrm{p}}}
\newcommand{\Pe}{\ensuremath{\mathrm{e}}}
\newcommand{\Pb}{\ensuremath{\mathrm{b}}}
\newcommand{\Pt}{\ensuremath{\mathrm{t}}}
\newcommand{\Pu}{\ensuremath{\mathrm{u}}}
\newcommand{\Pd}{\ensuremath{\mathrm{d}}}
\newcommand{\Ps}{\ensuremath{\mathrm{s}}}
\newcommand{\Pc}{\ensuremath{\mathrm{c}}}
\newcommand{\Pg}{\ensuremath{\mathrm{g}}}
  
\newcommand{\PW}{\ensuremath{\mathrm{W}}}
\newcommand{\PZ}{\ensuremath{\mathrm{Z}}}

\newcommand{\Mt}{\ensuremath{m_\Pt}\xspace}

\newcommand{\MW}{\ensuremath{M_\PW}\xspace}

\newcommand{\MZ}{\ensuremath{M_\PZ}\xspace}

\newcommand{\Gt}{\ensuremath{\Gamma_\Pt}\xspace}

\newcommand{\GZ}{\ensuremath{\Gamma_\PZ}\xspace}

\newcommand{\GW}{\ensuremath{\Gamma_\PW}\xspace}

\newcommand{\fb}{{\ensuremath\unskip\,\text{fb}}\xspace}

\newcommand{\GeV}{\ensuremath{\,\mathrm{GeV}}\xspace}
\newcommand{\TeV}{\ensuremath{\,\mathrm{TeV}}\xspace}

\newcommand{\alphas}{\ensuremath{\alpha_\mathrm{s}}\xspace}
\newcommand{\order}[1]{\ensuremath{\mathcal{O}{\left(#1\right)}}\xspace}

\newcommand{\newc}{\newcommand}
\newc{\bi}{\begin{itemize}}
\newc{\ei}{\end{itemize}}
\newc{\benu}{\begin{enumerate}}
\newc{\eenu}{\end{enumerate}}
\newc{\bc}{\begin{center}}
\newc{\ec}{\end{center}}
\newc{\bfig}{\begin{figure}}
\newc{\efig}{\end{figure}}
\newc{\qbar}{\bar{q}}
\newc{\go}{\tilde{g}}
\newc{\PB}{\mathrmsc{Powheg-Box}}

\newcommand{\recola}{{\sc Recola}\xspace}
\newcommand{\recolatwo}{{\sc Recola2}\xspace}

\newcommand{\mocanlo}{{\sc MoCaNLO}\xspace}
\newcommand{\collier}{{\sc Collier}\xspace}
\newcommand{\otter}{{\sc Otter}\xspace}

\newcommand{\rT}{{\mathrm{T}}}
\newcolumntype{.}{D{.}{.}{-1}}
\newcolumntype{d}[1]{D{.}{.}{#1}}

\colorlet{tableoverheadcolor}{gray!37.5}
\colorlet{tableheadcolor}{gray!25}
\colorlet{tablerowcolor}{gray!12.5}

\newlength{\width}
\newlength{\height}

\def\draftdate{\relax}
\def\mda{\relax}
\def\mua{\relax}
\def\mla{\relax}
\def\draft{
\def\thtystars{******************************}
\def\sixtystars{\thtystars\thtystars}
\typeout{}
\typeout{\sixtystars**}
\typeout{* Draft mode!
         For final version remove \protect\draft\space in source file *}
\typeout{\sixtystars**}
\typeout{}
\def\draftdate{\today}
\def\mua{\marginpar[\boldmath\hfil$\uparrow$]%
                   {\boldmath$\uparrow$\hfil}\color{black}%
                    \typeout{marginpar: $\uparrow$}\ignorespaces}
\def\mda{\color{red}\marginpar[\boldmath\hfil$\downarrow$]%
                   {\boldmath$\downarrow$\hfil}%
                    \typeout{marginpar: $\downarrow$}\ignorespaces}
\def\mla{\marginpar[\boldmath\hfil$\rightarrow$]%
                   {\boldmath$\leftarrow $\hfil}%
                    \typeout{marginpar: $\leftrightarrow$}\ignorespaces}
\def\Mua{\marginpar[\boldmath\hfil$\Uparrow$]%
                   {\boldmath$\Uparrow$\hfil}\color{black}%
                    \typeout{marginpar: $\uparrow$}\ignorespaces}
\def\Mda{\color{red}\marginpar[\boldmath\hfil$\Downarrow$]%
                   {\boldmath$\Downarrow$\hfil}%
                    \typeout{marginpar: $\downarrow$}\ignorespaces}
\def\Mla{\marginpar[\boldmath\hfil\textcolor{red}{$\Rightarrow$}]%
                   {\boldmath\textcolor{red}{$\Leftarrow $}\hfil}%
                    \typeout{marginpar: $\leftrightarrow$}\ignorespaces}
\overfullrule 5pt
\oddsidemargin 15mm
\marginparwidth 29mm
}


%% file: ttbb_PRD_rev.bbl
\providecommand{\href}[2]{#2}\begingroup\raggedright\begin{thebibliography}{10}

\bibitem{Beenakker:2001rj}
W.~Beenakker, et~al., {\it {Higgs radiation off top quarks at the Tevatron and
  the LHC}},  {\em Phys. Rev. Lett.} {\bf 87} (2001) 201805,
  [\href{http://arxiv.org/abs/hep-ph/0107081}{{\tt hep-ph/0107081}}].

\bibitem{Reina:2001sf}
L.~Reina and S.~Dawson, {\it {Next-to-leading order results for $t\bar{t}h$
  production at the Tevatron}},  {\em Phys. Rev. Lett.} {\bf 87} (2001) 201804,
  [\href{http://arxiv.org/abs/hep-ph/0107101}{{\tt hep-ph/0107101}}].

\bibitem{Beenakker:2002nc}
W.~Beenakker, et~al., {\it {NLO QCD corrections to $t\bar{t}H$ production in
  hadron collisions}},  {\em Nucl. Phys.} {\bf B653} (2003) 151--203,
  [\href{http://arxiv.org/abs/hep-ph/0211352}{{\tt hep-ph/0211352}}].

\bibitem{Dawson:2003zu}
S.~Dawson, C.~Jackson, L.~H. Orr, L.~Reina, and D.~Wackeroth, {\it {Associated
  Higgs production with top quarks at the large hadron collider: NLO QCD
  corrections}},  {\em Phys. Rev.} {\bf D68} (2003) 034022,
  [\href{http://arxiv.org/abs/hep-ph/0305087}{{\tt hep-ph/0305087}}].

\bibitem{Frederix:2011zi}
R.~Frederix, et~al., {\it {Scalar and pseudoscalar Higgs production in
  association with a top-antitop pair}},  {\em Phys. Lett.} {\bf B701} (2011)
  427--433, [\href{http://arxiv.org/abs/1104.5613}{{\tt arXiv:1104.5613}}].

\bibitem{Garzelli:2011vp}
M.~V. Garzelli, A.~Kardos, C.~G. Papadopoulos, and Z.~Trocsanyi, {\it {Standard
  Model Higgs boson production in association with a top anti-top pair at NLO
  with parton showering}},  {\em Europhys. Lett.} {\bf 96} (2011) 11001,
  [\href{http://arxiv.org/abs/1108.0387}{{\tt arXiv:1108.0387}}].

\bibitem{Hartanto:2015uka}
H.~B. Hartanto, B.~J{\"a}ger, L.~Reina, and D.~Wackeroth, {\it {Higgs boson
  production in association with top quarks in the POWHEG BOX}},  {\em Phys.
  Rev.} {\bf D91} (2015) 094003, [\href{http://arxiv.org/abs/1501.04498}{{\tt
  arXiv:1501.04498}}].

\bibitem{Denner:2015yca}
A.~Denner and R.~Feger, {\it {NLO QCD corrections to off-shell top-antitop
  production with leptonic decays in association with a Higgs boson at the
  LHC}},  {\em JHEP} {\bf 11} (2015) 209,
  [\href{http://arxiv.org/abs/1506.07448}{{\tt arXiv:1506.07448}}].

\bibitem{Kulesza:2015vda}
A.~Kulesza, L.~Motyka, T.~Stebel, and V.~Theeuwes, {\it {Soft gluon resummation
  for associated $t \bar{t} H$ production at the LHC}},  {\em JHEP} {\bf 03}
  (2016) 065, [\href{http://arxiv.org/abs/1509.02780}{{\tt arXiv:1509.02780}}].

\bibitem{Broggio:2015lya}
A.~Broggio, A.~Ferroglia, B.~D. Pecjak, A.~Signer, and L.~L. Yang, {\it
  {Associated production of a top pair and a Higgs boson beyond NLO}},  {\em
  JHEP} {\bf 03} (2016) 124, [\href{http://arxiv.org/abs/1510.01914}{{\tt
  arXiv:1510.01914}}].

\bibitem{Kulesza:2016vnq}
A.~Kulesza, L.~Motyka, T.~Stebel, and V.~Theeuwes, {\it {Soft gluon resummation
  at fixed invariant mass for associated $t\bar{t}H$ production at the LHC}},
  in {\em {4th Large Hadron Collider Physics Conference (LHCP 2016) Lund,
  Sweden, June 13-18, 2016}}, 2016.
\newblock \href{http://arxiv.org/abs/1609.01619}{{\tt arXiv:1609.01619}}.

\bibitem{Broggio:2016lfj}
A.~Broggio, A.~Ferroglia, B.~D. Pecjak, and L.~L. Yang, {\it {NNLL resummation
  for the associated production of a top pair and a Higgs boson at the LHC}},
  {\em JHEP} {\bf 02} (2017) 126, [\href{http://arxiv.org/abs/1611.00049}{{\tt
  arXiv:1611.00049}}].

\bibitem{Denner:2016wet}
A.~Denner, J.-N. Lang, M.~Pellen, and S.~Uccirati, {\it {Higgs production in
  association with off-shell top-antitop pairs at NLO EW and QCD at the LHC}},
  {\em JHEP} {\bf 02} (2017) 053, [\href{http://arxiv.org/abs/1612.07138}{{\tt
  arXiv:1612.07138}}].

\bibitem{Broggio:2019ewu}
A.~Broggio, et~al., {\it {Top-quark pair hadroproduction in association with a
  heavy boson at NLO+NNLL including EW corrections}},  {\em JHEP} {\bf 08}
  (2019) 039, [\href{http://arxiv.org/abs/1907.04343}{{\tt arXiv:1907.04343}}].

\bibitem{Kulesza:2020nfh}
A.~Kulesza, L.~Motyka, D.~Schwartl{\"a}nder, T.~Stebel, and V.~Theeuwes, {\it
  {Associated top quark pair production with a heavy boson: differential cross
  sections at NLO+NNLL accuracy}},  {\em Eur. Phys. J.} {\bf C80} (2020) 428,
  [\href{http://arxiv.org/abs/2001.03031}{{\tt arXiv:2001.03031}}].

\bibitem{Frixione:2014qaa}
S.~Frixione, V.~Hirschi, D.~Pagani, H.~S. Shao, and M.~Zaro, {\it {Weak
  corrections to Higgs hadroproduction in association with a top-quark pair}},
  {\em JHEP} {\bf 09} (2014) 065, [\href{http://arxiv.org/abs/1407.0823}{{\tt
  arXiv:1407.0823}}].

\bibitem{Yu:2014cka}
Y.~Zhang, W.-G. Ma, R.-Y. Zhang, C.~Chen, and L.~Guo, {\it {QCD NLO and EW NLO
  corrections to $t\bar{t}H$ production with top quark decays at hadron
  collider}},  {\em Phys. Lett.} {\bf B738} (2014) 1--5,
  [\href{http://arxiv.org/abs/1407.1110}{{\tt arXiv:1407.1110}}].

\bibitem{Frixione:2015zaa}
S.~Frixione, V.~Hirschi, D.~Pagani, H.~S. Shao, and M.~Zaro, {\it {Electroweak
  and QCD corrections to top-pair hadroproduction in association with heavy
  bosons}},  {\em JHEP} {\bf 06} (2015) 184,
  [\href{http://arxiv.org/abs/1504.03446}{{\tt arXiv:1504.03446}}].

\bibitem{Badger:2016bpw}
J.~R. Andersen et~al., {\it {Les Houches 2015: Physics at TeV Colliders
  Standard Model Working Group Report}},  in {\em {9th Les Houches Workshop on
  Physics at TeV Colliders (PhysTeV 2015) Les Houches, France, June 1-19,
  2015}}, 2016.
\newblock \href{http://arxiv.org/abs/1605.04692}{{\tt arXiv:1605.04692}}.

\bibitem{Bredenstein:2008zb}
A.~Bredenstein, A.~Denner, S.~Dittmaier, and S.~Pozzorini, {\it {NLO QCD
  corrections to $t\bar{t}b\bar{b}$ production at the LHC: 1. quark-antiquark
  annihilation}},  {\em JHEP} {\bf 08} (2008) 108,
  [\href{http://arxiv.org/abs/0807.1248}{{\tt arXiv:0807.1248}}].

\bibitem{Bredenstein:2009aj}
A.~Bredenstein, A.~Denner, S.~Dittmaier, and S.~Pozzorini, {\it {NLO QCD
  corrections to $pp\to t\bar{t}b\bar{b}+X$ at the LHC}},  {\em Phys. Rev.
  Lett.} {\bf 103} (2009) 012002, [\href{http://arxiv.org/abs/0905.0110}{{\tt
  arXiv:0905.0110}}].

\bibitem{Bevilacqua:2009zn}
G.~Bevilacqua, M.~Czakon, C.~Papadopoulos, R.~Pittau, and M.~Worek, {\it
  {Assault on the NLO Wishlist: $pp\to t\bar{t}b\bar{b}+X$}},  {\em JHEP} {\bf
  09} (2009) 109, [\href{http://arxiv.org/abs/0907.4723}{{\tt
  arXiv:0907.4723}}].

\bibitem{Bredenstein:2010rs}
A.~Bredenstein, A.~Denner, S.~Dittmaier, and S.~Pozzorini, {\it {NLO QCD
  corrections to $t\bar{t}b\bar{b}$ production at the LHC: 2. full hadronic
  results}},  {\em JHEP} {\bf 03} (2010) 021,
  [\href{http://arxiv.org/abs/1001.4006}{{\tt arXiv:1001.4006}}].

\bibitem{Cascioli:2013era}
F.~Cascioli, P.~Maierh{\"o}fer, N.~Moretti, S.~Pozzorini, and F.~Siegert, {\it
  {NLO matching for $t\bar t b \bar b$ production with massive $b$-quarks}},
  {\em Phys. Lett.} {\bf B734} (2014) 210--214,
  [\href{http://arxiv.org/abs/1309.5912}{{\tt arXiv:1309.5912}}].

\bibitem{Kardos:2013vxa}
A.~Kardos and Z.~Tr{\'o}cs{\'a}nyi, {\it {Hadroproduction of t anti-t pair with
  a b anti-b pair using PowHel}},  {\em J. Phys. G} {\bf 41} (2014) 075005,
  [\href{http://arxiv.org/abs/1303.6291}{{\tt arXiv:1303.6291}}].

\bibitem{Garzelli:2014aba}
M.~Garzelli, A.~Kardos, and Z.~Tr{\'o}cs{\'a}nyi, {\it {Hadroproduction of
  $t\bar{t}b\bar{b}$ final states at LHC: predictions at NLO accuracy matched
  with Parton Shower}},  {\em JHEP} {\bf 03} (2015) 083,
  [\href{http://arxiv.org/abs/1408.0266}{{\tt arXiv:1408.0266}}].

\bibitem{Bevilacqua:2017cru}
G.~Bevilacqua, M.~V. Garzelli, and A.~Kardos, {\it {$t\bar{t}b\bar{b}$
  hadroproduction with massive bottom quarks with PowHel}},
  \href{http://arxiv.org/abs/1709.06915}{{\tt arXiv:1709.06915}}.

\bibitem{Jezo:2018yaf}
T.~Je{\v z}o, J.~M. Lindert, N.~Moretti, and S.~Pozzorini, {\it {New NLOPS
  predictions for $t \bar{t} +b$-jet production at the LHC}},  {\em Eur. Phys.
  J.} {\bf C78} (2018) 502, [\href{http://arxiv.org/abs/1802.00426}{{\tt
  arXiv:1802.00426}}].

\bibitem{Buccioni:2019plc}
F.~Buccioni, S.~Kallweit, S.~Pozzorini, and M.~F. Zoller, {\it {NLO QCD
  predictions for $t\bar{t}b\bar{b}$ production in association with a light jet
  at the LHC}},  {\em JHEP} {\bf 12} (2019) 015,
  [\href{http://arxiv.org/abs/1907.13624}{{\tt arXiv:1907.13624}}].

\bibitem{deFlorian:2016spz}
{\bf LHC Higgs Cross Section Working Group} Collaboration, D.~de~Florian
  et~al., {\it {Handbook of LHC Higgs Cross Sections: 4. Deciphering the nature
  of the Higgs sector}},  (Geneva), CERN, 2016.
\newblock \href{http://arxiv.org/abs/1610.07922}{{\tt arXiv:1610.07922}}.
\newblock {CERN-2017-002-M}.

\bibitem{talkPozzoriniTOP2018}
S.~Pozzorini, ``{Theory progress on $t\bar{t}H(b\bar{b})$ background}.''
  \url{https://indico.cern.ch/event/690229/contributions/2979729/} (talk given
  at the 11th International Workshop on Top Quark Physics (TOP2018)).

\bibitem{talkSiegertZPW2020}
F.~Siegert, ``{ttH background systematics}.''
  \url{https://indico.cern.ch/event/857238/contributions/3643191/} (talk given
  at the Zurich Phenomenology Workshop 2020 (ZPW2020)).

\bibitem{Denner:2014wka}
A.~Denner, R.~Feger, and A.~Scharf, {\it {Irreducible background and
  interference effects for Higgs-boson production in association with a
  top-quark pair}},  {\em JHEP} {\bf 04} (2015) 008,
  [\href{http://arxiv.org/abs/1412.5290}{{\tt arXiv:1412.5290}}].

\bibitem{Bevilacqua:2019cvp}
G.~Bevilacqua, H.~B. Hartanto, M.~Kraus, T.~Weber, and M.~Worek, {\it {Towards
  constraining Dark Matter at the LHC: Higher order QCD predictions for
  $t\bar{t}+Z(Z\to \nu_\ell \bar{\nu}_\ell)$}},  {\em JHEP} {\bf 11} (2019)
  001, [\href{http://arxiv.org/abs/1907.09359}{{\tt arXiv:1907.09359}}].

\bibitem{Bevilacqua:2020pzy}
G.~Bevilacqua, H.-Y. Bi, H.~B. Hartanto, M.~Kraus, and M.~Worek, {\it {The
  simplest of them all: $t\bar{t} W^\pm$ at NLO accuracy in QCD}},  {\em JHEP}
  {\bf 08} (2020) 043, [\href{http://arxiv.org/abs/2005.09427}{{\tt
  arXiv:2005.09427}}].

\bibitem{Denner:2020hgg}
A.~Denner and G.~Pelliccioli, {\it {NLO QCD corrections to off-shell
  $\text{t}\bar{\text{t}}\text{W}^+$ production at the LHC}},  {\em JHEP} {\bf
  11} (2020) 069, [\href{http://arxiv.org/abs/2007.12089}{{\tt
  arXiv:2007.12089}}].

\bibitem{Anger:2017glm}
F.~R. Anger, F.~Febres~Cordero, H.~Ita, and V.~Sotnikov, {\it {NLO QCD
  predictions for $Wb\bar b$ production in association with up to three light
  jets at the LHC}},  {\em Phys. Rev.} {\bf D97} (2018) 036018,
  [\href{http://arxiv.org/abs/1712.05721}{{\tt arXiv:1712.05721}}].

\bibitem{Aaboud:2018eki}
{\bf ATLAS} Collaboration, M.~Aaboud et~al., {\it {Measurements of inclusive
  and differential fiducial cross-sections of $ t\overline{t} $ production with
  additional heavy-flavour jets in proton-proton collisions at $
  \sqrt{s}={}$13\,TeV with the ATLAS detector}},  {\em JHEP} {\bf 04} (2019)
  046, [\href{http://arxiv.org/abs/1811.12113}{{\tt arXiv:1811.12113}}].

\bibitem{Sirunyan:2020kga}
{\bf CMS} Collaboration, A.~M. Sirunyan et~al., {\it {Measurement of the cross
  section for $\text{t}\bar{\text{t}}$ production with additional jets and b
  jets in pp collisions at $\sqrt{s}={}$13\,TeV}},  {\em JHEP} {\bf 07} (2020)
  125, [\href{http://arxiv.org/abs/2003.06467}{{\tt arXiv:2003.06467}}].

\bibitem{Aaboud:2018urx}
{\bf ATLAS} Collaboration, M.~Aaboud et~al., {\it {Observation of Higgs boson
  production in association with a top quark pair at the LHC with the ATLAS
  detector}},  {\em Phys. Lett. B} {\bf 784} (2018) 173--191,
  [\href{http://arxiv.org/abs/1806.00425}{{\tt arXiv:1806.00425}}].

\bibitem{Sirunyan:2018hoz}
{\bf CMS} Collaboration, A.~M. Sirunyan et~al., {\it {Observation of
  $t\overline{t}$H production}},  {\em Phys. Rev. Lett.} {\bf 120} (2018)
  231801, [\href{http://arxiv.org/abs/1804.02610}{{\tt arXiv:1804.02610}}].

\bibitem{Aaboud:2017rss}
{\bf ATLAS} Collaboration, M.~Aaboud et~al., {\it {Search for the standard
  model Higgs boson produced in association with top quarks and decaying into a
  $b\bar{b}$ pair in $pp$ collisions at $\sqrt{s}={}$13\,TeV with the ATLAS
  detector}},  {\em Phys. Rev. D} {\bf 97} (2018) 072016,
  [\href{http://arxiv.org/abs/1712.08895}{{\tt arXiv:1712.08895}}].

\bibitem{Sirunyan:2018mvw}
{\bf CMS} Collaboration, A.~M. Sirunyan et~al., {\it {Search for
  ${t}\overline{t}{H} $ production in the ${H}\to{b}\overline{{b}} $ decay
  channel with leptonic ${t}\overline{t} $ decays in proton-proton collisions
  at $\sqrt{s}={}$13\,TeV}},  {\em JHEP} {\bf 03} (2019) 026,
  [\href{http://arxiv.org/abs/1804.03682}{{\tt arXiv:1804.03682}}].

\bibitem{Actis:2012qn}
S.~Actis et~al., {\it {Recursive generation of one-loop amplitudes in the
  Standard Model}},  {\em JHEP} {\bf 04} (2013) 037,
  [\href{http://arxiv.org/abs/1211.6316}{{\tt arXiv:1211.6316}}].

\bibitem{Actis:2016mpe}
S.~Actis et~al., {\it {RECOLA: REcursive Computation of One-Loop Amplitudes}},
  {\em Comput. Phys. Commun.} {\bf 214} (2017) 140--173,
  [\href{http://arxiv.org/abs/1605.01090}{{\tt arXiv:1605.01090}}].

\bibitem{Denner:2017vms}
A.~Denner, J.-N. Lang, and S.~Uccirati, {\it {NLO electroweak corrections in
  extended Higgs Sectors with RECOLA2}},  {\em JHEP} {\bf 07} (2017) 087,
  [\href{http://arxiv.org/abs/1705.06053}{{\tt arXiv:1705.06053}}].

\bibitem{Denner:2017wsf}
A.~Denner, J.-N. Lang, and S.~Uccirati, {\it {Recola2: REcursive Computation of
  One-Loop Amplitudes 2}},  {\em Comput. Phys. Commun.} {\bf 224} (2018)
  346--361, [\href{http://arxiv.org/abs/1711.07388}{{\tt arXiv:1711.07388}}].

\bibitem{otter:2020}
F.~Buccioni, J.-N. Lang, and S.~Pozzorini, {\it {OTTER: On-The-fly TEnsor
  Reduction}},  {\em To be published}.

\bibitem{Denner:2014gla}
A.~Denner, S.~Dittmaier, and L.~Hofer, {\it {COLLIER - A fortran-library for
  one-loop integrals}},  {\em PoS} {\bf LL2014} (2014) 071,
  [\href{http://arxiv.org/abs/1407.0087}{{\tt arXiv:1407.0087}}].

\bibitem{Denner:2016kdg}
A.~Denner, S.~Dittmaier, and L.~Hofer, {\it {{\sc Collier}: a fortran-based
  Complex One-Loop LIbrary in Extended Regularizations}},  {\em Comput. Phys.
  Commun.} {\bf 212} (2017) 220--238,
  [\href{http://arxiv.org/abs/1604.06792}{{\tt arXiv:1604.06792}}].

\bibitem{Denner:2000bj}
A.~Denner, S.~Dittmaier, M.~Roth, and D.~Wackeroth, {\it {Electroweak radiative
  corrections to $e^+ e^-\to W W\to 4\,$fermions in double pole approximation:
  The RACOONWW approach}},  {\em Nucl. Phys.} {\bf B587} (2000) 67--117,
  [\href{http://arxiv.org/abs/hep-ph/0006307}{{\tt hep-ph/0006307}}].

\bibitem{Denner:2016jyo}
A.~Denner and M.~Pellen, {\it {NLO electroweak corrections to off-shell
  top-antitop production with leptonic decays at the LHC}},  {\em JHEP} {\bf
  08} (2016) 155, [\href{http://arxiv.org/abs/1607.05571}{{\tt
  arXiv:1607.05571}}].

\bibitem{Denner:2017kzu}
A.~Denner and M.~Pellen, {\it {Off-shell production of top-antitop pairs in the
  lepton+jets channel at NLO QCD}},  {\em JHEP} {\bf 02} (2018) 013,
  [\href{http://arxiv.org/abs/1711.10359}{{\tt arXiv:1711.10359}}].

\bibitem{Denner:1999gp}
A.~Denner et~al., {\it {Predictions for all processes $e^+ e^-\to
  4\,$fermions${}+ \gamma$}},  {\em Nucl. Phys.} {\bf B560} (1999) 33--65,
  [\href{http://arxiv.org/abs/hep-ph/9904472}{{\tt hep-ph/9904472}}].

\bibitem{Denner:2005fg}
A.~Denner et~al., {\it {Electroweak corrections to charged-current $e^+ e^-\to
  4\,$fermion processes: Technical details and further results}},  {\em Nucl.
  Phys.} {\bf B724} (2005) 247--294,
  [\href{http://arxiv.org/abs/hep-ph/0505042}{{\tt hep-ph/0505042}}].

\bibitem{Denner:2019vbn}
A.~Denner and S.~Dittmaier, {\it {Electroweak Radiative Corrections for
  Collider Physics}},  {\em Phys. Rept.} {\bf 864} (2020) 1--163,
  [\href{http://arxiv.org/abs/1912.06823}{{\tt arXiv:1912.06823}}].

\bibitem{Denner:1997ia}
A.~Denner, S.~Dittmaier, and M.~Roth, {\it {Non-factorizable photonic
  corrections to $e^+ e^-\to W W \to\,$four fermions}},  {\em Nucl. Phys.} {\bf
  B519} (1998) 39--84, [\href{http://arxiv.org/abs/hep-ph/9710521}{{\tt
  hep-ph/9710521}}].

\bibitem{Accomando:2004de}
E.~Accomando, A.~Denner, and A.~Kaiser, {\it {Logarithmic electroweak
  corrections to gauge-boson pair production at the LHC}},  {\em Nucl. Phys. B}
  {\bf 706} (2005) 325--371, [\href{http://arxiv.org/abs/hep-ph/0409247}{{\tt
  hep-ph/0409247}}].

\bibitem{Dittmaier:2015bfe}
S.~Dittmaier and C.~Schwan, {\it {Non-factorizable photonic corrections to
  resonant production and decay of many unstable particles}},  {\em Eur. Phys.
  J.} {\bf C76} (2016) 144, [\href{http://arxiv.org/abs/1511.01698}{{\tt
  arXiv:1511.01698}}].

\bibitem{Biedermann:2016yds}
B.~Biedermann, A.~Denner, and M.~Pellen, {\it {Large electroweak corrections to
  vector-boson scattering at the Large Hadron Collider}},  {\em Phys. Rev.
  Lett.} {\bf 118} (2017) 261801, [\href{http://arxiv.org/abs/1611.02951}{{\tt
  arXiv:1611.02951}}].

\bibitem{Denner:2020bcz}
A.~Denner and G.~Pelliccioli, {\it {Polarized electroweak bosons in ${\bf
  \text{W}^+\text{W}^-}$ production at the LHC including NLO QCD effects}},
  {\em JHEP} {\bf 09} (2020) 164, [\href{http://arxiv.org/abs/2006.14867}{{\tt
  arXiv:2006.14867}}].

\bibitem{Nagy:1998bb}
Z.~Nagy and Z.~Trocsanyi, {\it {Next-to-leading order calculation of four-jet
  observables in electron-positron annihilation}},  {\em Phys. Rev.} {\bf D59}
  (1999) 014020, [\href{http://arxiv.org/abs/hep-ph/9806317}{{\tt
  hep-ph/9806317}}]. [Erratum: Phys. Rev. {\bf D62} (2000) 099902].

\bibitem{Berends:1994pv}
F.~A. Berends, R.~Pittau, and R.~Kleiss, {\it {All electroweak four fermion
  processes in electron-positron collisions}},  {\em Nucl. Phys.} {\bf B424}
  (1994) 308--342, [\href{http://arxiv.org/abs/hep-ph/9404313}{{\tt
  hep-ph/9404313}}].

\bibitem{Dittmaier:2002ap}
S.~Dittmaier and M.~Roth, {\it {LUSIFER: A LUcid approach to six FERmion
  production}},  {\em Nucl. Phys.} {\bf B642} (2002) 307--343,
  [\href{http://arxiv.org/abs/hep-ph/0206070}{{\tt hep-ph/0206070}}].

\bibitem{Buccioni:2017yxi}
F.~Buccioni, S.~Pozzorini, and M.~Zoller, {\it {On-the-fly reduction of open
  loops}},  {\em Eur. Phys. J. C} {\bf 78} (2018) 70,
  [\href{http://arxiv.org/abs/1710.11452}{{\tt arXiv:1710.11452}}].

\bibitem{Buccioni:2019sur}
F.~Buccioni, et~al., {\it {OpenLoops 2}},  {\em Eur. Phys. J. C} {\bf 79}
  (2019) 866, [\href{http://arxiv.org/abs/1907.13071}{{\tt arXiv:1907.13071}}].

\bibitem{vanHameren:2010cp}
A.~van Hameren, {\it {OneLOop: For the evaluation of one-loop scalar
  functions}},  {\em Comput. Phys. Commun.} {\bf 182} (2011) 2427--2438,
  [\href{http://arxiv.org/abs/1007.4716}{{\tt arXiv:1007.4716}}].

\bibitem{Catani:1996vz}
S.~Catani and M.~H. Seymour, {\it {A general algorithm for calculating jet
  cross-sections in NLO QCD}},  {\em Nucl. Phys.} {\bf B485} (1997) 291--419,
  [\href{http://arxiv.org/abs/hep-ph/9605323}{{\tt hep-ph/9605323}}]. [Erratum:
  Nucl. Phys. {\bf B510} (1998) 503].

\bibitem{Tanabashi:2018oca}
{\bf ParticleDataGroup} Collaboration, M.~Tanabashi et~al., {\it {Review of
  Particle Physics}},  {\em Phys. Rev.} {\bf D98} (2018) 030001.

\bibitem{Bardin:1988xt}
D.~{\relax Yu}. Bardin, A.~Leike, T.~Riemann, and M.~Sachwitz, {\it
  {Energy-dependent width effects in ${e}^+ {e}^-$-annihilation near the
  Z-boson pole}},  {\em Phys. Lett.} {\bf B206} (1988) 539--542.

\bibitem{Jezabek:1988iv}
M.~Je\.{z}abek and J.~H. K{\"u}hn, {\it {QCD Corrections to Semileptonic Decays
  of Heavy Quarks}},  {\em Nucl. Phys.} {\bf B314} (1989) 1--6.

\bibitem{Basso:2015gca}
L.~Basso, S.~Dittmaier, A.~Huss, and L.~Oggero, {\it {Techniques for the
  treatment of IR divergences in decay processes at NLO and application to the
  top-quark decay}},  {\em Eur. Phys. J.} {\bf C76} (2016) 56,
  [\href{http://arxiv.org/abs/1507.04676}{{\tt arXiv:1507.04676}}].

\bibitem{Ball:2017nwa}
{\bf NNPDF} Collaboration, R.~D. Ball et~al., {\it {Parton distributions from
  high-precision collider data}},  {\em Eur. Phys. J. C} {\bf 77} (2017) 663,
  [\href{http://arxiv.org/abs/1706.00428}{{\tt arXiv:1706.00428}}].

\bibitem{Andersen:2014efa}
J.~R. Andersen et~al., {\it {Les Houches 2013: Physics at TeV Colliders:
  Standard Model Working Group Report}},  in {\em {8th Les Houches Workshop on
  Physics at TeV Colliders (PhysTeV 2013) Les Houches, France, June 3-21,
  2013}}, 2014.
\newblock \href{http://arxiv.org/abs/1405.1067}{{\tt arXiv:1405.1067}}.

\bibitem{Buckley:2014ana}
A.~Buckley, et~al., {\it {LHAPDF6: parton density access in the LHC precision
  era}},  {\em Eur. Phys. J.} {\bf C75} (2015) 132,
  [\href{http://arxiv.org/abs/1412.7420}{{\tt arXiv:1412.7420}}].

\bibitem{Cacciari:2008gp}
M.~Cacciari, G.~P. Salam, and G.~Soyez, {\it {The anti-$k_t$ jet clustering
  algorithm}},  {\em JHEP} {\bf 04} (2008) 063,
  [\href{http://arxiv.org/abs/0802.1189}{{\tt arXiv:0802.1189}}].

\bibitem{Bevilacqua:2021cit}
G.~Bevilacqua, et~al., {\it {$t\bar{t}b\bar{b}$ at the LHC: On the size of
  corrections and $b$-jet definitions}},
  \href{http://arxiv.org/abs/2105.08404}{{\tt arXiv:2105.08404}}.

\bibitem{Cascioli:2013gfa}
F.~Cascioli et~al., {\it {Precise Higgs-background predictions: merging NLO QCD
  and squared quark-loop corrections to four-lepton + 0,1 jet production}},
  {\em JHEP} {\bf 01} (2014) 046, [\href{http://arxiv.org/abs/1309.0500}{{\tt
  arXiv:1309.0500}}].

\bibitem{Bevilacqua:2010qb}
G.~Bevilacqua, M.~Czakon, A.~van Hameren, C.~G. Papadopoulos, and M.~Worek,
  {\it {Complete off-shell effects in top quark pair hadroproduction with
  leptonic decay at next-to-leading order}},  {\em JHEP} {\bf 02} (2011) 083,
  [\href{http://arxiv.org/abs/1012.4230}{{\tt arXiv:1012.4230}}].

\bibitem{Denner:2012yc}
A.~Denner, S.~Dittmaier, S.~Kallweit, and S.~Pozzorini, {\it {NLO QCD
  corrections to off-shell top-antitop production with leptonic decays at
  hadron colliders}},  {\em JHEP} {\bf 10} (2012) 110,
  [\href{http://arxiv.org/abs/1207.5018}{{\tt arXiv:1207.5018}}].

\bibitem{AlcarazMaestre:2012vp}
{\bf SM, NLO MULTILEG Working Group, SM MC Working Group} Collaboration,
  J.~Alcaraz~Maestre et~al., {\it {The SM and NLO Multileg and SM MC Working
  Groups: Summary Report}},  in {\em {7th Les Houches Workshop on Physics at
  TeV Colliders}}, 3, 2012.
\newblock \href{http://arxiv.org/abs/1203.6803}{{\tt arXiv:1203.6803}}.

\end{thebibliography}\endgroup
